\documentclass[pra,twocolumn,preprintnumbers,amsmath,amssymb]{revtex4-1}
\usepackage{amsmath}
\usepackage{amssymb}
\usepackage{mathrsfs}
\usepackage{graphicx}
\usepackage{dcolumn}
\usepackage{bm}
\usepackage{color}
\usepackage{hhline}
\usepackage{rotating}

\begin{document}

\title{Theory of high-resolution tunneling spin transport on a magnetic skyrmion}

\author{Kriszti\'an Palot\'as$^{1,2}$}
\email{palotas@phy.bme.hu}
\author{Levente R\'ozsa$^{3}$}
\author{L\'aszl\'o Szunyogh$^{4,5}$}

\affiliation{1 Slovak Academy of Sciences, Institute of Physics, SK-84511 Bratislava, Slovakia\\
2 University of Szeged, MTA-SZTE Reaction Kinetics and Surface Chemistry Research Group, H-6720 Szeged, Hungary\\
3 University of Hamburg, Department of Physics, D-20355 Hamburg, Germany\\
4 Budapest University of Technology and Economics, Department of Theoretical Physics, H-1111 Budapest, Hungary\\
5 Budapest University of Technology and Economics, MTA-BME Condensed Matter Research Group, H-1111 Budapest, Hungary}

\date{\today}

\begin{abstract}

Tunneling spin transport characteristics of a magnetic skyrmion are described theoretically in magnetic scanning tunneling microscopy (STM). The spin-polarized charge current in STM (SP-STM) and tunneling spin transport vector quantities, the longitudinal spin current and the spin transfer torque, are calculated in high spatial resolution within the same theoretical framework. A connection between the conventional charge current SP-STM image contrasts and the magnitudes of the spin transport vectors is demonstrated that enables the estimation of tunneling spin transport properties based on experimentally measured SP-STM images. A considerable tunability of the spin transport vectors by the involved spin polarizations is also highlighted. These possibilities and the combined theory of tunneling charge and vector spin transport pave the way for gaining deep insight into electric-current-induced tunneling spin transport properties in SP-STM and to the related dynamics of complex magnetic textures at surfaces.

\end{abstract}

\maketitle

\section{Introduction}

Due to the electron's intrinsic properties, electronic transport includes both charge and spin transfer. Understanding the mechanisms and increasing the efficiency of current-induced magnetization switching (CIMS) have a large impact on the ongoing development of magnetic spintronic devices. Consequently, CIMS attracted considerable research interest in metallic spin valves and magnetic tunnel junctions (MTJs) in the recent years \cite{ralph08}. So far, theoretical research of spin transport partly focused on the calculation of the tunneling spin transfer torque (STT) in symmetric or asymmetric planar MTJs. The theoretical approaches cover free-electron models \cite{slonczewski96,wilczynski08,manchon08}, a transfer Hamiltonian method \cite{slonczewski05}, tight-binding-based transport models \cite{theodonis06,kalitsov09,kalitsov13,pauyac14,useinov15}, and scattering methods \cite{xiao08}, partly combined with first-principles techniques \cite{heiliger08,jia11}. The first direct measurement of the out-of-plane and in-plane components of the STT vector in a planar MTJ was provided by Sankey {\it et al.} \cite{sankey08}. The tunability of the magnetotransport properties in material combinations of current interest has also been predicted recently \cite{ukpong18}.

Magnetic skyrmions in thin films are real-space spin textures possessing a topological charge \cite{leonov16}, and in most of the studied cases they show a preferred chirality due to the emergence of the antisymmetric Dzyaloshinsky-Moriya exchange interaction \cite{Dzyaloshinsky,Moriya} in the magnetic layer with broken inversion symmetry\cite{romming13,Yu,Tokunaga,moreau16,woo16}. On the other hand, it has been demonstrated that frustrated Heisenberg exchange interactions may also lead to the stabilization of localized skyrmionic spin configurations with different topologies \cite{Okubo,Leonov,Lin,rozsa-sk3}. A range of technological applications of isolated skyrmions has been proposed for magnetic data storage and transfer \cite{fert13,nagaosa13,zhang15,wiesendanger16natrevmat,fert17natrevmat}. The finite temperature stability of skyrmions has been investigated both experimentally \cite{moreau16,boulle16,woo16} and theoretically \cite{hagemeister15,rozsa-sk1,lobanov16}, and the annihilation of skyrmions has recently been extensively studied by using minimum energy path calculations \cite{lobanov16,bessarab15,bessarab17,rohart17,stosic17}.

Magnetic skyrmions in ultrathin films can conveniently be imaged by using spin-polarized scanning tunneling microscopy (SP-STM) \cite{bergmann14,romming15prl,palotas17prb}, and first-principles calculations have proven crucial in understanding the formation of these spin structures through determining microscopic magnetic interactions \cite{heinze11,dupe14,simon14,rozsa-sk2}. Over the last two decades, the powerful capabilities of SP-STM have been demonstrated for obtaining local information on spin-polarized tunneling charge transport properties of a great variety of magnetic surfaces \cite{wiesendanger09review}. High-resolution SP-STM is expected to further contribute, e.g., to the proper understanding of the interaction of skyrmions with defects that is currently an important challenge \cite{hanneken16,stosic17_2}. Recently, there has been a growing interest in getting insight into local STT properties in high spatial resolution in asymmetric MTJs. Naturally, the tunneling STT, which is concomitantly present with the charge current, contributes to local CIMS effects \cite{krause11}. Two recent examples on the atomic scale include Refs.\ \onlinecite{romming13} and \onlinecite{hsu17elec}, where local current pulses with opposite voltage polarities have been used with an STM tip to create and annihilate individual magnetic skyrmions in different thin film systems. However, the detailed microscopic mechanism of the local tunneling STT and another spin transport component, the longitudinal spin current (LSC) \cite{manchon16chapter}, is not clarified in these processes yet.

In this paper the generalization of high-resolution STM charge transport theories \cite{hofer03rmp} is proposed to include vector spin transport in STM junctions going beyond the assumption of collinear magnetic structure for both the surface and the tip used in Ref.\ \onlinecite{palotas16prb}. A combined electron tunneling theory for the charge and vector spin transport in magnetic STM junctions is presented considering complex noncollinear surface magnetic structures within the three-dimensional (3D) Wentzel-Kramers-Brillouin (WKB) electron tunneling framework. The theory is implemented in the 3D-WKB-STM code \cite{palotas14fop}, which is an established and efficient method for the simulation of spin-polarized scanning tunneling microscopy \cite{palotas11stm,palotas12orb,palotas13contrast,mandi14fe,nita14} and spectroscopy \cite{palotas11sts,palotas12sts} with an enhanced parameter space for modeling tip geometries \cite{mandi13tiprot,mandi14rothopg,mandi15tipstat}. Besides the spin-polarized charge current above complex magnetic textures in SP-STM \cite{palotas17prb,palotas11stm,palotas13contrast,wortmann01,heinze06}, the proposed method allows for the high-resolution calculation of tunneling spin transport vector quantities, the STT and the LSC. The connections between the SP-STM image contrasts of the charge current and the magnitudes of the STT as well as the LSC are highlighted, and it is concluded that estimations on the tunneling spin transport properties can be made based on experimental SP-STM images. Moreover, a direct relationship between the scalar spin polarization and the ratio of the out-of-plane and the in-plane STT components is given, and it is proposed that the measurement of the STT vector components would enable the determination of the spin polarizations separately for the sample surface and the tip. The influence of these spin polarizations on the tunneling spin transport is also analyzed, and a considerable tunability of the spin transport properties is pointed out. Finally, theoretical considerations are reported on the measurement of the STT and the LSC vectors.

The paper is organized as follows. The 3D-WKB combined tunneling electron charge and vector spin transport theory in magnetic STM considering noncollinear magnetic surfaces is presented in detail in Sec.\ \ref{sec_theory}, where the dominating atomic contributions for the interpretation of our results are described in Sec.\ \ref{sec_atom}. Relationships between the charge current and the magnitudes of the LSC and the STT are reported in Sec.\ \ref{sec_ilscstt}, and tunneling parameters and visualization remarks are given in Sec.\ \ref{sec_param}. The vector spin transport (LSC and STT) properties of a skyrmion in relation to the spin-polarized charge current are analyzed in Sec.\ \ref{sec_chsp}, and the effect of the spin polarizations is investigated in Sec.\ \ref{sec_spinpol}. Summary and conclusions are found in Sec.\ \ref{sec_conc}, and spin transport vector measurement considerations are given in Appendix.

\section{Theory and methods}
\label{sec_th}

\subsection{3D-WKB theory of combined electron charge and spin tunneling}
\label{sec_theory}

The 3D-WKB theoretical model for the combined tunneling charge and vector spin transport in magnetic STM above complex noncollinear magnetic surfaces is based on the recently introduced electron tunneling model through STM junctions built up from collinear magnetic surface and tip \cite{palotas16prb}. Note that in the present paper we do not restrict the formalism to fixed spin quantization axes of the sample and the tip, as this would describe collinear magnets. Instead, the spatial dependence of the local atomic spin quantization axes of the noncollinear magnetic surface is allowed. For simplicity, we omit the orbital dependence \cite{palotas12orb,mandi14fe} of the electronic structure taken into account in Ref.\ \onlinecite{palotas16prb}, and the low bias limit is employed.

Consider the following normalized (dimensionless) density matrices in spin space for the sample surface ($S$) atom labeled by "$a$" and the tip ($T$) apex atom at the corresponding Fermi levels $E_F^S$ and $E_F^T$ \cite{hofer03magn,palotas11stm,palotas16prb}:
\begin{subequations}
\begin{align}
\underline{\underline{\rho}}_{S}^a(E_F^S)&=\underline{\underline{I}}+\mathbf{P}_{S}^a(E_F^S)\cdot\underline{\underline{\bm{\sigma}}},\\
\underline{\underline{\rho}}_{T}(E_F^T)&=\underline{\underline{I}}+\mathbf{P}_{T}(E_F^T)\cdot\underline{\underline{\bm{\sigma}}},
\end{align}
\end{subequations}
where $\underline{\underline{I}}$ is the $2\times 2$ unit matrix and $\underline{\underline{\bm{\sigma}}}=\left(\underline{\underline{\sigma}}_x,\underline{\underline{\sigma}}_y,\underline{\underline{\sigma}}_z\right)$ is a vector composed of the Pauli matrices. $\mathbf{P}_{S}^a(E_F^S)=\mathbf{m}_{S}^a(E_F^S)/n_{S}^a(E_F^S)$ is the spin polarization vector of the atom-projected density of states (PDOS) of the sample surface atom "$a$", where $\mathbf{m}_{S}^a(E_F^S)$ and $n_{S}^a(E_F^S)$ denote the magnetization (vector) and charge (scalar) character of the corresponding PDOS at $E_F^S$, respectively. Similarly, $\mathbf{P}_{T}(E_F^T)=\mathbf{m}_{T}(E_F^T)/n_{T}(E_F^T)$, where $\mathbf{P}_{T}(E_F^T)$, $\mathbf{m}_{T}(E_F^T)$ and $n_{T}(E_F^T)$ denote the spin polarization PDOS vector, magnetization PDOS vector, and charge PDOS of the tip apex atom at $E_F^T$. These quantities can be calculated by using {\it{ab initio}} electronic structure methods \cite{palotas11stm}, or can be treated as parameters as in the present paper.

The coupled dimensionless transport quantities, the scalar charge conductance ($\tilde{I}$) and the generalized vector spin torkance ($\tilde{\mathbf{T}}$), of the electron tunneling between the sample surface atom "$a$" and the tip apex atom at the common Fermi levels $E_F^S=E_F^T=E_F$ (zero bias limit) can be represented by traces in spin space, and  they depend on the tunneling direction tip$\rightarrow$sample ($T\rightarrow S$) or sample$\rightarrow$tip ($S\rightarrow T$) as
\begin{subequations}
\begin{align}
\tilde{I}^{a,T\rightarrow S}(E_F)&=\frac{1}{2}\mathrm{Tr}\left(\underline{\underline{\rho}}_{T}(E_F)\underline{\underline{I}}\underline{\underline{\rho}}_{S}^a(E_F)\right)\nonumber\\
=\tilde{I}^{a,S\rightarrow T}(E_F)&=\frac{1}{2}\mathrm{Tr}\left(\underline{\underline{\rho}}_{S}^a(E_F)\underline{\underline{I}}\underline{\underline{\rho}}_{T}(E_F)\right)\nonumber\\
=\tilde{I}^{a}(E_F)&=1+\mathbf{P}_{S}^a(E_F)\cdot\mathbf{P}_{T}(E_F),\label{Eq_chsp1}\\
\tilde{\mathbf{T}}^{a,T\rightarrow S}(E_F)&=\frac{1}{2}\mathrm{Tr}\left(\underline{\underline{\rho}}_{T}(E_F)\underline{\underline{\bm{\sigma}}}\underline{\underline{\rho}}_{S}^a(E_F)\right)\nonumber\\
&=\mathbf{P}_{S}^a(E_F)+\mathbf{P}_{T}(E_F)\nonumber\\
&+i\mathbf{P}_{S}^a(E_F)\times\mathbf{P}_{T}(E_F),\label{Eq_chsp2}\\
\tilde{\mathbf{T}}^{a,S\rightarrow T}(E_F)&=\frac{1}{2}\mathrm{Tr}\left(\underline{\underline{\rho}}_{S}^a(E_F)\underline{\underline{\bm{\sigma}}}\underline{\underline{\rho}}_{T}(E_F)\right)\nonumber\\
&=\mathbf{P}_{S}^a(E_F)+\mathbf{P}_{T}(E_F)\nonumber\\
&-i\mathbf{P}_{S}^a(E_F)\times\mathbf{P}_{T}(E_F).\label{Eq_chsp3}
\end{align}
\end{subequations}
Since $\mathbf{P}_{S}^a(E_F)\cdot\mathbf{P}_{T}(E_F)=\left[P_{S}^a(E_F)\mathbf{s}_{S}^a\right]\cdot\left[P_{T}(E_F)\mathbf{s}_{T}\right]=P_{S}^a(E_F)P_{T}(E_F)\cos\phi_a$ with $\phi_a$ the angle between the classical spin unit vectors of the sample surface atom "$a$" ($\mathbf{s}_{S}^a=\mathbf{S}_S^a/|\mathbf{S}_S^a|$) and the tip apex ($\mathbf{s}_{T}=\mathbf{S}_T/|\mathbf{S}_T|$), where $\mathbf{S}_S^a$ and $\mathbf{S}_T$ are the corresponding spin moments, the charge conductance formula is formally equivalent to the spin-polarized Tersoff-Hamann model \cite{tersoff83,tersoff85,wortmann01,heinze06}:
\begin{equation}
\tilde{I}^{a}(E_F)=1+P_{S}^a(E_F)P_{T}(E_F)\cos\phi_a.
\end{equation}
Here, the first and second term is the non-magnetic and magnetic (spin-polarized) part of the dimensionless charge conductance at zero bias voltage, respectively.

The in-plane and out-of-plane components of the generalized vector spin torkance formula are obtained as the real and imaginary parts of Eqs.~(\ref{Eq_chsp2}) and (\ref{Eq_chsp3}), respectively:
\begin{subequations}
\begin{align}
&\tilde{\mathbf{T}}^{a,in-pl.}(E_F)=\mathbf{P}_{S}^a(E_F)+\mathbf{P}_{T}(E_F)\nonumber\\
&=\mathrm{Re}\left\{\tilde{\mathbf{T}}^{a,T\rightarrow S}(E_F)\right\}=\mathrm{Re}\left\{\tilde{\mathbf{T}}^{a,S\rightarrow T}(E_F)\right\},\label{Eq_T1}\\
&\tilde{\mathbf{T}}^{a,out-pl.}(E_F)=\mathbf{P}_S^a(E_F)\times\mathbf{P}_T(E_F)\nonumber\\
&=\mathrm{Im}\left\{\tilde{\mathbf{T}}^{a,T\rightarrow S}(E_F)\right\}=-\mathrm{Im}\left\{\tilde{\mathbf{T}}^{a,S\rightarrow T}(E_F)\right\}.\label{Eq_T2}
\end{align}
\end{subequations}
Here, the in-plane component $\tilde{\mathbf{T}}^{a,in-pl.}$ lies in the plane spanned by the $\mathbf{P}_{S}^a(E_F)$ and $\mathbf{P}_{T}(E_F)$ (or $\mathbf{s}_S^a$ and $\mathbf{s}_T$) vectors, depending on the considered lattice site $a$. However, the direction of $\tilde{\mathbf{T}}^{a,in-pl.}$ is not necessarily perpendicular to the spin moment ($\mathbf{S}_T$ or $\mathbf{S}_S^a$) on which the generalized torkance is acting. Therefore, $\tilde{\mathbf{T}}^{in-pl.}$ can be decomposed into a dimensionless longitudinal spin conductance part denoted by $L$, parallel to the spin moment directions: $\mathbf{s}_T$ or local $\mathbf{s}_S^a$ unit vectors; and into a dimensionless spin torkance part denoted by $\parallel$, perpendicular to the spin moment directions \cite{xiao08}. The decomposition reads
$\tilde{\mathbf{T}}^{a,in-pl.}(E_F)=\tilde{\mathbf{T}}^{a,j,L}(E_F)+\tilde{\mathbf{T}}^{a,j,\parallel}(E_F)$ with $j\in\{T,S\}$, where $\parallel$ denotes that this in-plane (Slonczewski or damping-like) torkance vector is in the $\mathbf{S}_S^a$--$\mathbf{S}_T$ (or $\mathbf{s}_S^a$--$\mathbf{s}_T$) plane. The corresponding components are
\begin{subequations}
\begin{align}
\tilde{\mathbf{T}}^{a,T,L}(E_F)&=\left[\tilde{\mathbf{T}}^{a,in-pl.}(E_F)\cdot\mathbf{s}_T\right]\mathbf{s}_T\nonumber\\
&=\left(P_S^a(E_F)\cos\phi_a+P_T(E_F)\right)\mathbf{s}_T,\label{Eq_Tpar1}\\
\tilde{\mathbf{T}}^{a,T,\parallel}(E_F)&=\tilde{\mathbf{T}}^{a,in-pl.}(E_F)-\tilde{\mathbf{T}}^{a,T,L}(E_F)\nonumber\\
&=P_S^a(E_F)\left(\mathbf{s}_S^a-\mathbf{s}_T\cos\phi_a\right)\nonumber\\
&=P_S^a(E_F)\mathbf{s}_T\times\left(\mathbf{s}_S^a\times\mathbf{s}_T\right),\label{Eq_Tpar2}\\
\tilde{\mathbf{T}}^{a,S,L}(E_F)&=\left[\tilde{\mathbf{T}}^{a,in-pl.}(E_F)\cdot\mathbf{s}_S^a\right]\mathbf{s}_S^a\nonumber\\
&=\left(P_S^a(E_F)+P_T(E_F)\cos\phi_a\right)\mathbf{s}_S^a,\label{Eq_Tpar3}\\
\tilde{\mathbf{T}}^{a,S,\parallel}(E_F)&=\tilde{\mathbf{T}}^{a,in-pl.}(E_F)-\tilde{\mathbf{T}}^{a,S,L}(E_F)\nonumber\\
&=P_T(E_F)\left(\mathbf{s}_T-\mathbf{s}_S^a\cos\phi_a\right)\nonumber\\
&=P_T(E_F)\mathbf{s}_S^a\times\left(\mathbf{s}_T\times\mathbf{s}_S^a\right).\label{Eq_Tpar4}
\end{align}
\end{subequations}
Here, the $T$ or $S$ indices denote the tip or sample side of the tunnel junction, on which the longitudinal spin conductance and the in-plane torkance are acting. As it is clear from Eq.~(\ref{Eq_T1}), these quantities are independent of the actual current flow direction $T\rightarrow S$ or $S\rightarrow T$; for experimental evidence for the in-plane torkance, see, e.g., Fig.\ 3a in Ref.\ \onlinecite{sankey08}.

The out-of-plane component in Eq.~(\ref{Eq_T2}), $\tilde{\mathbf{T}}^{a,out-pl.}$, is perpendicular to the plane spanned by the $\mathbf{P}_{S}^a(E_F)$ and $\mathbf{P}_{T}(E_F)$ (or $\mathbf{s}_S^a$ and $\mathbf{s}_T$) vectors, and the dimensionless out-of-plane (field-like) torkance vector can be identified as \cite{xiao08}
\begin{eqnarray}
\tilde{\mathbf{T}}^{a,\perp}(E_F)&=&\tilde{\mathbf{T}}^{a,out-pl.}(E_F)=P_{S}^a(E_F)P_{T}(E_F)\mathbf{s}_S^a\times\mathbf{s}_T\nonumber\\
&=&\tilde{\mathbf{T}}^{a,T\rightarrow S,\perp}(E_F)=-\tilde{\mathbf{T}}^{a,S\rightarrow T,\perp}(E_F).
\label{Eq_Tper}
\end{eqnarray}
According to Eq.~(\ref{Eq_T2}), the out-of-plane torkance vector changes sign by reversing the current flow direction; for experimental evidence see, e.g., Fig.\ 3a in Ref.\ \onlinecite{sankey08}.
Note that throughout the paper the out-of-plane torkance and torque is purely current-induced, resulting from electron tunneling through the vacuum barrier, and the equilibrium torque \cite{theodonis06} is not taken into account. Combining Eqs.~(\ref{Eq_T1}), (\ref{Eq_T2}), (\ref{Eq_Tpar2}), (\ref{Eq_Tpar4}) and (\ref{Eq_Tper}), the following dimensionless torkance vectors are obtained:
\begin{subequations}
\begin{align}
\tilde{\mathbf{T}}^{a,T\rightarrow S,T}(E_F)&=\tilde{\mathbf{T}}^{a,T,\parallel}(E_F)+\tilde{\mathbf{T}}^{a,\perp}(E_F),\\
\tilde{\mathbf{T}}^{a,S\rightarrow T,T}(E_F)&=\tilde{\mathbf{T}}^{a,T,\parallel}(E_F)-\tilde{\mathbf{T}}^{a,\perp}(E_F),\\
\tilde{\mathbf{T}}^{a,T\rightarrow S,S}(E_F)&=\tilde{\mathbf{T}}^{a,S,\parallel}(E_F)+\tilde{\mathbf{T}}^{a,\perp}(E_F),\\
\tilde{\mathbf{T}}^{a,S\rightarrow T,S}(E_F)&=\tilde{\mathbf{T}}^{a,S,\parallel}(E_F)-\tilde{\mathbf{T}}^{a,\perp}(E_F),
\end{align}
\end{subequations}
where $\tilde{\mathbf{T}}^{a,T\rightarrow S,T}$ and $\tilde{\mathbf{T}}^{a,S\rightarrow T,T}$ act on the spin moment of the tip apex atom at $T\rightarrow S$ and $S\rightarrow T$ tunneling, respectively. Similarly, $\tilde{\mathbf{T}}^{a,T\rightarrow S,S}$ and $\tilde{\mathbf{T}}^{a,S\rightarrow T,S}$ act on the spin moment of the sample surface atom "$a$" at the indicated tunneling directions.

Assuming elastic electron tunneling, the signed charge current ($I_s$) can be obtained from the charge conductance within the 3D-WKB theory by the superposition of atomic contributions from the sample surface (sum over "$a$") \cite{palotas11stm} at the tip apex position $\mathbf{R}_{T}$ and at bias voltage $V$ in the low bias limit as
\begin{equation}
I_s(\mathbf{R}_{T},V)=\frac{e^2}{2\pi\hbar}V\sum_a h(\mathbf{R}_{T}-\mathbf{R}_{a})\tilde{I}^a(E_F).
\label{Eq_current}
\end{equation}
The transmission function,
\begin{equation}
h(\mathbf{r})=\exp\left[-\sqrt{8m\Phi/\hbar^2}|\mathbf{r}|\right]
\label{Eq_Transmission}
\end{equation}
(with the electron's mass $m$, the reduced Planck constant $\hbar$, and the effective work function $\Phi$), depends on the relative position of the tip apex atom and the sample surface atom "$a$" ($\mathbf{R}_{T}-\mathbf{R}_a$), and in the transmission all electron states are considered as exponentially decaying spherical states \cite{tersoff83,tersoff85,heinze06}, neglecting orbital dependence \cite{palotas12orb}.

The signed longitudinal spin current (LSC) vector ($\mathbf{T}_s^{L}$) and the signed spin transfer torque (STT) vector components ($\mathbf{T}_s^{\parallel}$ and $\mathbf{T}_s^{\perp}$) acting on the spin moment of the tip apex atom can be calculated from the corresponding longitudinal spin conductance and torkance components, respectively, as
\begin{subequations}
\begin{align}
\mathbf{T}_s^{TL}(\mathbf{R}_{T},V)&=eV\sum_a h(\mathbf{R}_{T}-\mathbf{R}_{a})\tilde{\mathbf{T}}^{a,T,L}(E_F),\label{Eq_torque_Ta}\\
\mathbf{T}_s^{T\parallel}(\mathbf{R}_{T},V)&=eV\sum_a h(\mathbf{R}_{T}-\mathbf{R}_{a})\tilde{\mathbf{T}}^{a,T,\parallel}(E_F),\label{Eq_torque_Tb}\\
\mathbf{T}_s^{T\rightarrow S,T\perp}(\mathbf{R}_{T},V)&=eV\sum_a h(\mathbf{R}_{T}-\mathbf{R}_{a})\tilde{\mathbf{T}}^{a,T\rightarrow S,\perp}(E_F),\label{Eq_torque_Tc}\\
\mathbf{T}_s^{S\rightarrow T,T\perp}(\mathbf{R}_{T},V)&=eV\sum_a h(\mathbf{R}_{T}-\mathbf{R}_{a})\tilde{\mathbf{T}}^{a,S\rightarrow T,\perp}(E_F).\label{Eq_torque_Td}
\end{align}
\end{subequations}
Similarly, the signed LSC vector and the signed STT vector components acting on the spin moment of surface atom "$a$" at position $\mathbf{R}_a$ are obtained as
\begin{subequations}
\begin{align}
\mathbf{T}_s^{aSL}(\mathbf{R}_a,V)&=eVh(\mathbf{R}_{T}-\mathbf{R}_{a})\tilde{\mathbf{T}}^{a,S,L}(E_F),\label{Eq_torque_Sa}\\
\mathbf{T}_s^{aS\parallel}(\mathbf{R}_a,V)&=eVh(\mathbf{R}_{T}-\mathbf{R}_{a})\tilde{\mathbf{T}}^{a,S,\parallel}(E_F),\label{Eq_torque_Sb}\\
\mathbf{T}_s^{a,T\rightarrow S,S\perp}(\mathbf{R}_a,V)&=eVh(\mathbf{R}_{T}-\mathbf{R}_{a})\tilde{\mathbf{T}}^{a,T\rightarrow S,\perp}(E_F),\label{Eq_torque_Sc}\\
\mathbf{T}_s^{a,S\rightarrow T,S\perp}(\mathbf{R}_a,V)&=eVh(\mathbf{R}_{T}-\mathbf{R}_{a})\tilde{\mathbf{T}}^{a,S\rightarrow T,\perp}(E_F).\label{Eq_torque_Sd}
\end{align}
\end{subequations}
In Eqs.~(\ref{Eq_current}), (\ref{Eq_torque_Ta})-(\ref{Eq_torque_Td}) and (\ref{Eq_torque_Sa})-(\ref{Eq_torque_Sd}) the signed charge current and the signed LSC and STT vectors have the correct dimensions by multiplying the dimensionless atomic charge conductance, spin conductance and torkance contributions with proper factors, i.e., $I_s=\frac{e^2}{2\pi\hbar}V\tilde{I}$ and $\mathbf{T}_s^{L/\parallel/\perp}=eV\tilde{\mathbf{T}}^{L/\parallel/\perp}$, where $e$ is the elementary charge and $V$ is the bias voltage.

Taking the sign convention of the bias voltage into account, i.e., $V>0$ at $T\rightarrow S$ and $V<0$ at $S\rightarrow T$ tunneling, we find that the LSC and the in-plane STT vectors actually change sign and the out-of-plane STT vectors do not change sign by reversing the bias polarity, thus the direction of the current flow. These are clearly seen in Eqs.~(\ref{Eq_torque_Ta})-(\ref{Eq_torque_Td}) and (\ref{Eq_torque_Sa})-(\ref{Eq_torque_Sd}): in $\mathbf{T}_s^{L/\parallel}$ $V$ changes sign and according to Eq.~(\ref{Eq_T1}) $\tilde{\mathbf{T}}^{L/\parallel}$ do not, and in $\mathbf{T}_s^{\perp}$ both $V$ and $\tilde{\mathbf{T}}^{\perp}$ change sign, the latter according to Eq.~(\ref{Eq_Tper}). These findings are in agreement with previous spin transport interpretations \cite{ralph08}. The absolute charge current is independent of the tunneling direction, and can be calculated as
\begin{eqnarray}
&&I(\mathbf{R}_{T},V)\nonumber\\
&=&I_s(\mathbf{R}_{T},V>0)=-I_s(\mathbf{R}_{T},V<0)\nonumber\\
&=&\frac{e^2}{2\pi\hbar}|V|\sum_a h(\mathbf{R}_{T}-\mathbf{R}_{a})(1+P_{S}^a(E_F)P_{T}(E_F)\cos\phi_a).\nonumber\\
\label{Eq_current1}
\end{eqnarray}
Similarly, the spin transport vectors acting on the spin moment of the tip apex atom are obtained from Eqs.~(\ref{Eq_torque_Ta})-(\ref{Eq_torque_Td}) as
\begin{subequations}
\begin{align}
&\mathbf{T}^{TL}(\mathbf{R}_{T},V)\nonumber\\
&=\mathbf{T}_s^{TL}(\mathbf{R}_{T},V>0)=-\mathbf{T}_s^{TL}(\mathbf{R}_{T},V<0)\nonumber\\
&=e|V|\sum_a h(\mathbf{R}_{T}-\mathbf{R}_{a})\left(P_S^a(E_F)\cos\phi_a+P_T(E_F)\right)\mathbf{s}_T,\label{Eq_torque_T1a}\\
&\mathbf{T}^{T\parallel}(\mathbf{R}_{T},V)\nonumber\\
&=\mathbf{T}_s^{T\parallel}(\mathbf{R}_{T},V>0)=-\mathbf{T}_s^{T\parallel}(\mathbf{R}_{T},V<0)\nonumber\\
&=e|V|\sum_a h(\mathbf{R}_{T}-\mathbf{R}_{a})P_S^a(E_F)\mathbf{s}_T\times\left(\mathbf{s}_S^a\times\mathbf{s}_T\right),\label{Eq_torque_T1b}\\
&\mathbf{T}^{\perp}(\mathbf{R}_{T},V)\nonumber\\
&=\mathbf{T}_s^{T\rightarrow S,T\perp}(\mathbf{R}_{T},V>0)=\mathbf{T}_s^{S\rightarrow T,T\perp}(\mathbf{R}_{T},V<0)\nonumber\\
&=e|V|\sum_a h(\mathbf{R}_{T}-\mathbf{R}_{a})P_{S}^a(E_F)P_{T}(E_F)\mathbf{s}_S^a\times\mathbf{s}_T,\label{Eq_torque_T1c}\\
&\mathbf{T}^{T\rightarrow S,T}(\mathbf{R}_{T},V>0)=\mathbf{T}^{\perp}(\mathbf{R}_{T},V)+\mathbf{T}^{T\parallel}(\mathbf{R}_{T},V),\label{Eq_torque_T1d}\\
&\mathbf{T}^{S\rightarrow T,T}(\mathbf{R}_{T},V<0)=\mathbf{T}^{\perp}(\mathbf{R}_{T},V)-\mathbf{T}^{T\parallel}(\mathbf{R}_{T},V),\label{Eq_torque_T1e}
\end{align}
\end{subequations}
and the spin transport vectors acting on the spin moment of surface atom "$a$" are obtained from Eqs.~(\ref{Eq_torque_Sa})-(\ref{Eq_torque_Sd}) as
\begin{subequations}
\begin{align}
&\mathbf{T}^{aSL}(\mathbf{R}_a,V)\nonumber\\
&=\mathbf{T}_s^{aSL}(\mathbf{R}_a,V>0)=-\mathbf{T}_s^{aSL}(\mathbf{R}_a,V<0)\nonumber\\
&=e|V|h(\mathbf{R}_{T}-\mathbf{R}_{a})\left(P_S^a(E_F)+P_T(E_F)\cos\phi_a\right)\mathbf{s}_S^a,\label{Eq_torque_S1a}\\
&\mathbf{T}^{aS\parallel}(\mathbf{R}_a,V)\nonumber\\
&=\mathbf{T}_s^{aS\parallel}(\mathbf{R}_a,V>0)=-\mathbf{T}_s^{aS\parallel}(\mathbf{R}_a,V<0)\nonumber\\
&=e|V|h(\mathbf{R}_{T}-\mathbf{R}_{a})P_T(E_F)\mathbf{s}_S^a\times\left(\mathbf{s}_T\times\mathbf{s}_S^a\right),\label{Eq_torque_S1b}\\
&\mathbf{T}^{a\perp}(\mathbf{R}_a,V)\nonumber\\
&=\mathbf{T}_s^{a,T\rightarrow S,S\perp}(\mathbf{R}_a,V>0)=\mathbf{T}_s^{a,S\rightarrow T,S\perp}(\mathbf{R}_a,V<0)\nonumber\\
&=e|V|h(\mathbf{R}_{T}-\mathbf{R}_{a})P_{S}^a(E_F)P_{T}(E_F)\mathbf{s}_S^a\times\mathbf{s}_T,\label{Eq_torque_S1c}\\
&\mathbf{T}^{a,T\rightarrow S,S}(\mathbf{R}_a,V>0)=\mathbf{T}^{a\perp}(\mathbf{R}_a,V)+\mathbf{T}^{aS\parallel}(\mathbf{R}_a,V),\label{Eq_torque_S1d}\\
&\mathbf{T}^{a,S\rightarrow T,S}(\mathbf{R}_a,V<0)=\mathbf{T}^{a\perp}(\mathbf{R}_a,V)-\mathbf{T}^{aS\parallel}(\mathbf{R}_a,V),\label{Eq_torque_S1e}
\end{align}
\end{subequations}
where the notations $\mathbf{T}^{TL}$ and $\mathbf{T}^{aSL}$ for the LSC vectors, $\mathbf{T}^{T\parallel}$ and $\mathbf{T}^{aS\parallel}$ for the in-plane STT vectors, and $\mathbf{T}^{\perp}=\sum_a\mathbf{T}^{a\perp}$ for the out-of-plane STT vectors have been introduced.

The LSC and the STT acting on the sample for the scanning tip at position $\mathbf{R}_{T}$ are defined as the sum of the vector spin transport quantities acting on the different surface atoms "$a$",
\begin{subequations}
\begin{align}
&\mathbf{T}^{SL}(\mathbf{R}_{T},V)\nonumber\\
&=e|V|\sum_a h(\mathbf{R}_{T}-\mathbf{R}_{a})\left(P_S^a(E_F)+P_T(E_F)\cos\phi_a\right)\mathbf{s}_S^a,\label{Eq_torque_S2a}\\
&\mathbf{T}^{S\parallel}(\mathbf{R}_{T},V)\nonumber\\
&=e|V|\sum_a h(\mathbf{R}_{T}-\mathbf{R}_{a})P_T(E_F)\mathbf{s}_S^a\times\left(\mathbf{s}_T\times\mathbf{s}_S^a\right),\label{Eq_torque_S2b}\\
&\mathbf{T}^{\perp}(\mathbf{R}_{T},V)\nonumber\\
&=e|V|\sum_a h(\mathbf{R}_{T}-\mathbf{R}_{a})P_{S}^a(E_F)P_{T}(E_F)\mathbf{s}_S^a\times\mathbf{s}_T,\label{Eq_torque_S2c}\\
&\mathbf{T}^{T\rightarrow S,S}(\mathbf{R}_{T},V>0)=\mathbf{T}^{\perp}(\mathbf{R}_{T},V)+\mathbf{T}^{S\parallel}(\mathbf{R}_{T},V),\label{Eq_torque_S2d}\\
&\mathbf{T}^{S\rightarrow T,S}(\mathbf{R}_{T},V<0)=\mathbf{T}^{\perp}(\mathbf{R}_{T},V)-\mathbf{T}^{S\parallel}(\mathbf{R}_{T},V).\label{Eq_torque_S2e}
\end{align}
\end{subequations}
As discussed in Sec.\ \ref{sec_atom}, these values are dominated by the contributions coming from the closest surface atom $A$ below the STM tip. Note also the equivalence of Eqs.~(\ref{Eq_torque_T1c}) and (\ref{Eq_torque_S2c}).

Throughout the paper the reported electronic charge and vector spin transport quantities correspond to a scanning tip at position $\mathbf{R}_T$ and to Eqs.~(\ref{Eq_current1}), (\ref{Eq_torque_T1a})-(\ref{Eq_torque_T1e}) and (\ref{Eq_torque_S2a})-(\ref{Eq_torque_S2e}). These are key results of the present paper. The transport components in a simplified fashion are summarized below for a better overview:
\begin{subequations}
\begin{align}
I(\mathbf{R}_{T})&\propto\sum_a h(\mathbf{R}_{T}-\mathbf{R}_a)(1+P_{S}P_{T}\cos\phi_a),\label{Eq_1a}\\
\mathbf{T}^{TL}(\mathbf{R}_{T})&\propto\sum_a h(\mathbf{R}_{T}-\mathbf{R}_a)(P_{T}+P_{S}\cos\phi_a)\mathbf{s}_T,\label{Eq_1b}\\
\mathbf{T}^{SL}(\mathbf{R}_{T})&\propto\sum_a h(\mathbf{R}_{T}-\mathbf{R}_a)(P_{S}+P_{T}\cos\phi_a)\mathbf{s}_S^a,\label{Eq_1c}\\
\mathbf{T}^{T\parallel}(\mathbf{R}_{T})&\propto\sum_a h(\mathbf{R}_{T}-\mathbf{R}_a)P_{S}\mathbf{s}_T\times(\mathbf{s}_S^a\times\mathbf{s}_T),\label{Eq_1d}\\
\mathbf{T}^{S\parallel}(\mathbf{R}_{T})&\propto\sum_a h(\mathbf{R}_{T}-\mathbf{R}_a)P_{T}\mathbf{s}_S^a\times(\mathbf{s}_T\times\mathbf{s}_S^a),\label{Eq_1e}\\
\mathbf{T}^{\perp}(\mathbf{R}_{T})&\propto\sum_a h(\mathbf{R}_{T}-\mathbf{R}_a)P_{S}P_{T}\mathbf{s}_S^a\times\mathbf{s}_T,\label{Eq_1f}
\end{align}
\end{subequations}
and the total STT vectors are
\begin{subequations}
\begin{align}
\mathbf{T}^{T\rightarrow S,T}(\mathbf{R}_{T})&=\mathbf{T}^{\perp}(\mathbf{R}_{T})+\mathbf{T}^{T\parallel}(\mathbf{R}_{T}),\label{Eq_2a}\\
\mathbf{T}^{S\rightarrow T,T}(\mathbf{R}_{T})&=\mathbf{T}^{\perp}(\mathbf{R}_{T})-\mathbf{T}^{T\parallel}(\mathbf{R}_{T}),\label{Eq_2b}\\
\mathbf{T}^{T\rightarrow S,S}(\mathbf{R}_{T})&=\mathbf{T}^{\perp}(\mathbf{R}_{T})+\mathbf{T}^{S\parallel}(\mathbf{R}_{T}),\label{Eq_2c}\\
\mathbf{T}^{S\rightarrow T,S}(\mathbf{R}_{T})&=\mathbf{T}^{\perp}(\mathbf{R}_{T})-\mathbf{T}^{S\parallel}(\mathbf{R}_{T}).\label{Eq_2d}
\end{align}
\end{subequations}
Here, it was assumed that $P_S^a(E_F)=P_S$ for all surface atoms, and the notation $P_T(E_F)=P_T$ was used for simplicity. Note that the effective spin polarization ($P_{\textrm{eff}}=P_SP_T$) only enters the charge current (Eq.~(\ref{Eq_1a})) and the out-of-plane torque (Eq.~(\ref{Eq_1f})) expressions, and not the longitudinal spin current or the in-plane torque. This means that $P_{\textrm{eff}}$ is not sufficient to characterize the spin polarization of the magnetic tunnel junction concerning spin transport quantities, and $P_S$ and $P_T$ are independent parameters in our model. The effect of these spin polarizations is investigated on the tunneling spin transport properties of a magnetic skyrmion in Sec.\ \ref{sec_spinpol}.

\subsection{Dominating atomic contributions}
\label{sec_atom}

Due to the exponential decay of the tunneling transmission in Eq.\ (\ref{Eq_Transmission}), the atomic sums in the tunneling charge and vector spin transport quantities in Eqs.\ (\ref{Eq_current}), (\ref{Eq_torque_Ta})-(\ref{Eq_torque_Td}), (\ref{Eq_current1}), (\ref{Eq_torque_T1a})-(\ref{Eq_torque_T1c}), (\ref{Eq_torque_S2a})-(\ref{Eq_torque_S2c}) and (\ref{Eq_1a})-(\ref{Eq_1f}) are convergent. They are dominated, and thus can be approximated by the sum of the contributions from the closest surface atoms below the tip position $\mathbf{R}_{T}$. Such a set of surface atoms can be denoted by $\mathcal{A}(\mathbf{R}_{T})$. The selection of surface atoms "$a$" in the set of $\mathcal{A}(\mathbf{R}_{T})$ depends on a properly chosen convergence criterion \cite{palotas11sts}. Consequently, Eqs.\ (\ref{Eq_current}), (\ref{Eq_torque_Ta})-(\ref{Eq_torque_Td}), (\ref{Eq_current1}), (\ref{Eq_torque_T1a})-(\ref{Eq_torque_T1c}), (\ref{Eq_torque_S2a})-(\ref{Eq_torque_S2c}) and (\ref{Eq_1a})-(\ref{Eq_1f}) can be interpreted as $\mathbf{R}_{T}$-dependent weighted averages over the set of surface atoms $\mathcal{A}(\mathbf{R}_{T})$, e.g., $I_s(\mathbf{R}_{T},V)\approx\frac{e^2}{2\pi\hbar}V\tilde{I}^{\mathcal{A}(\mathbf{R}_{T})}(E_F)$ with $\tilde{I}^{\mathcal{A}(\mathbf{R}_{T})}(E_F)=\sum_{a\in\mathcal{A}(\mathbf{R}_{T})}h(\mathbf{R}_{T}-\mathbf{R}_{a})\tilde{I}^a(E_F)$. Although in the paper the sum over "$a$" is performed for all sample atoms in the simulated area (for more details see Sec.\ \ref{sec_param}), for the interpretation of the results the dominating contribution is considered to come from the closest surface atom below the STM tip, which is denoted by $A$ and characterized by the spin unit vector $\mathbf{s}_S^A$. Clearly, all quantities denoted by $A$ depend on the lateral position of the tip, just as above: $A(\mathbf{R}_{T})$. Following this, the tunneling electron charge and vector spin transport components can be written as
\begin{subequations}
\begin{align}
I(\mathbf{R}_{T},V)&\propto1+P_{S}P_{T}\cos\phi_A,\label{Eq_vectors1}\\
\mathbf{T}^{TL}(\mathbf{R}_{T},V)&\propto\left(P_{S}\cos\phi_A+P_{T}\right)\mathbf{s}_T,\label{Eq_vectors2}\\
\mathbf{T}^{SL}(\mathbf{R}_{T},V)\approx\mathbf{T}^{ASL}(\mathbf{R}_A,V)&\propto\left(P_{T}\cos\phi_A+P_{S}\right)\mathbf{s}_S^A,\label{Eq_vectors3}\\
\mathbf{T}^{T\parallel}(\mathbf{R}_{T},V)&\propto P_{S}\mathbf{s}_T\times(\mathbf{s}_S^A\times\mathbf{s}_T),\label{Eq_vectors4}\\
\mathbf{T}^{S\parallel}(\mathbf{R}_{T},V)\approx\mathbf{T}^{AS\parallel}(\mathbf{R}_A,V)&\propto P_{T}\mathbf{s}_S^A\times(\mathbf{s}_T\times\mathbf{s}_S^A),\label{Eq_vectors5}\\
\mathbf{T}^{\perp}(\mathbf{R}_{T},V)\approx\mathbf{T}^{A\perp}(\mathbf{R}_A,V)&\propto P_{S}P_{T}\mathbf{s}_S^A\times\mathbf{s}_T,\label{Eq_vectors6}
\end{align}
\end{subequations}
with $\cos\phi_A=\mathbf{s}_S^A\cdot\mathbf{s}_T$, and the magnitudes of the vector spin transport quantities are
\begin{subequations}
\begin{align}
|\mathbf{T}^{TL}(\mathbf{R}_{T},V)|&\propto\left|P_{S}\cos\phi_A+P_{T}\right|,\label{Eq_magnitudes1}\\
|\mathbf{T}^{SL}(\mathbf{R}_{T},V)|\approx|\mathbf{T}^{ASL}(\mathbf{R}_A,V)|&\propto\left|P_{T}\cos\phi_A+P_{S}\right|,\label{Eq_magnitudes2}\\
|\mathbf{T}^{T\parallel}(\mathbf{R}_{T},V)|&\propto|P_{S}\sin\phi_A|,\label{Eq_magnitudes3}\\
|\mathbf{T}^{S\parallel}(\mathbf{R}_{T},V)|\approx|\mathbf{T}^{AS\parallel}(\mathbf{R}_A,V)|&\propto|P_{T}\sin\phi_A|,\label{Eq_magnitudes4}\\
|\mathbf{T}^{\perp}(\mathbf{R}_{T},V)|\approx|\mathbf{T}^{A\perp}(\mathbf{R}_A,V)|&\propto|P_{S}P_{T}\sin\phi_A|,\label{Eq_magnitudes5}\\
|\mathbf{T}^{T}(\mathbf{R}_{T},V)|&\propto|P_{S}\sin\phi_A|\sqrt{1+P_{T}^2},\label{Eq_magnitudes6}\\
|\mathbf{T}^{S}(\mathbf{R}_{T},V)|\approx|\mathbf{T}^{AS}(\mathbf{R}_A,V)|&\propto|P_{T}\sin\phi_A|\sqrt{1+P_{S}^2},\label{Eq_magnitudes7}
\end{align}
\end{subequations}
where $|\mathbf{T}^{T}|=|\mathbf{T}^{T\rightarrow S,T}|=|\mathbf{T}^{S\rightarrow T,T}|$, $|\mathbf{T}^{S}|=|\mathbf{T}^{T\rightarrow S,S}|=|\mathbf{T}^{S\rightarrow T,S}|$, and $|\mathbf{T}^{AS}|=|\mathbf{T}^{A,T\rightarrow S,S}|=|\mathbf{T}^{A,S\rightarrow T,S}|$. Note that both STT components ($\parallel$ and $\perp$), and thus the STT obey the expected $\sin\phi_A$-dependence.

\subsection{Connections between the charge current and the magnitudes of the LSC and the STT}
\label{sec_ilscstt}

Following the previous section for the dominating atomic contributions to the tunneling electron transport properties, simple relationships between the charge current and the spin transport magnitudes can be derived. Let us assume that the charge current can be measured at opposite tip magnetization directions $\mathbf{s}_T$ and $-\mathbf{s}_T$. This results in the charge currents $I(\mathbf{s}_T)\propto1+P_{S}P_{T}\cos\phi_A$ and $I(-\mathbf{s}_T)\propto1-P_{S}P_{T}\cos\phi_A$, from which the spin-polarized contribution \cite{palotas13contrast} to the current (also known as magnetic asymmetry \cite{palotas12sts}, $A_I$) can be expressed as
\begin{equation}
A_I=P_{S}P_{T}\cos\phi_A=\frac{I(\mathbf{s}_T)-I(-\mathbf{s}_T)}{I(\mathbf{s}_T)+I(-\mathbf{s}_T)}.
\label{Eq_AI}
\end{equation}
This quantity takes values between $-1$ and $+1$, and can directly be obtained in experiments in the differential magnetic mode \cite{wulfhekel10} of SP-STM, or in simulations employing, e.g., the above-described 3D-WKB model. Note that similar magnetic asymmetry quantities can be defined for the longitudinal spin currents, see Appendix. Following the above, $\cos\phi_A=A_I/(P_{S}P_{T})$, $|\sin\phi_A|=\sqrt{P_{S}^2P_{T}^2-A_I^2}/|P_{S}P_{T}|$, and the charge and spin transport magnitudes assuming an $\mathbf{s}_T$ tip magnetization direction can be written as
\begin{subequations}
\begin{align}
I&\propto1+A_I,\label{Eq_A1}\\
|\mathbf{T}^{TL}|&\propto\left|P_{T}+A_I/P_{T}\right|,\label{Eq_A2}\\
|\mathbf{T}^{SL}|&\propto\left|P_{S}+A_I/P_{S}\right|,\label{Eq_A3}\\
|\mathbf{T}^{T\parallel}|&\propto\sqrt{P_{S}^2P_{T}^2-A_I^2}/|P_{T}|,\label{Eq_A4}\\
|\mathbf{T}^{S\parallel}|&\propto\sqrt{P_{S}^2P_{T}^2-A_I^2}/|P_{S}|,\label{Eq_A5}\\
|\mathbf{T}^{\perp}|&\propto\sqrt{P_{S}^2P_{T}^2-A_I^2},\label{Eq_A6}\\
|\mathbf{T}^{T}|&\propto\sqrt{P_{S}^2P_{T}^2-A_I^2}\sqrt{1+1/P_{T}^2},\label{Eq_A7}\\
|\mathbf{T}^{S}|&\propto\sqrt{P_{S}^2P_{T}^2-A_I^2}\sqrt{1+1/P_{S}^2}.\label{Eq_A8}
\end{align}
\end{subequations}
These equations establish a direct connection between the magnitudes of the spin transport components and the charge current through the spin-polarized contribution of the latter, $A_I$.

Considering the lateral $(x,y)$-dependence of a scanning tip, the approximate $\phi_A$ can be replaced by an effective $\phi(x,y)$ in a continuum description, which exactly reproduces the electron transport components in Eqs.\ (\ref{Eq_1a})-(\ref{Eq_1f}), and the above derivation also applies in a point by point fashion for high-resolution images using the $A_I(x,y)$ expression based on Eq.\ (\ref{Eq_AI}): $A_I(x,y)=P_{S}P_{T}\cos\phi(x,y)=[I(x,y,\mathbf{s}_T)-I(x,y,-\mathbf{s}_T)]/[I(x,y,\mathbf{s}_T)+I(x,y,-\mathbf{s}_T)]$.

\subsection{Model parameters and visualization remarks}
\label{sec_param}

In the following we report on computational parameters used in the electron charge and spin tunneling model. The spin structure of the noncollinear magnetic surface is an input parameter of the 3D-WKB-STM code. The considered spin structure of a skyrmion was taken from Ref.\ \onlinecite{rozsa-sk3}, where it was relaxed on a single-layer triangular lattice with $C_{3v}$ crystallographic symmetry containing $128\times128=16384$ lattice sites using spin dynamics simulations. The underlying magnetic interaction parameters of Fe in the (Pt$_{0.95}$Ir$_{0.05}$)/Fe/Pd(111) ultrathin film system were obtained from {\it{ab initio}} calculations \cite{rozsa-sk2}. We note that the antisymmetric Dzyaloshinsky-Moriya and the frustrated Heisenberg magnetic exchange interactions are concomitantly present in this ultrathin magnetic film \cite{rozsa-sk3}.

The absolute bias voltage is set to $|V|=1.5$ meV and the effective work function to $\Phi=5$ eV. Motivated by the reported electronic structure of a recent work \cite{crum15}, $P_S=-0.5$ together with $P_T=-0.8$ are chosen in Sec.\ \ref{sec_chsp}, thus resulting in $P_{\textrm{eff}}=+0.4$, which value was also considered in previous works discussing SP-STM characteristics of skyrmionic spin textures with different topologies \cite{palotas17prb,dupe16stm}. For investigating the effect of the spin polarizations on the spin transport, the combinations of the sets $P_S\in\{-0.5,+0.5\}$ and $P_T\in\{-0.8,-0.4,+0.4,+0.8\}$ are considered in Sec.\ \ref{sec_spinpol}.

Tunneling charge and spin transport quantities are calculated in a scan area of 7.5 nm $\times$ 6 nm. SP-STM images of the charge current are shown in constant-current mode using a white-brown-black color palette corresponding to maximum-medium-minimum apparent heights. Employing the reported parameters, the current value of $I=10^{-4}$ nA of the constant-current contours corresponds to about 6 \AA\;minimal tip-sample distance and corrugation values between 30 and 40 pm \cite{palotas17prb}. The spin transport (STT and LSC) quantities (vectors and scalar magnitudes) are given in constant-height mode at 6 \AA\;tip-sample distance. The magnitudes of the STT and LSC are shown using a red-green-blue color palette corresponding to maximum-medium-minimum values of the individual images. While the STT and LSC vectors are calculated in the same high lateral resolution as the charge current and the magnitudes of the STT and the LSC (1 \AA\; resolution for all), for visualization reasons the lateral resolution of the vector spin transport quantities is set to 5 \AA.

\section{Results and discussion}
\label{sec_res}

\subsection{Tunneling charge and spin transport properties of a skyrmion}
\label{sec_chsp}

\begin{figure}[t]
\begin{tabular}{c}
\includegraphics[width=1.0\columnwidth,angle=0]{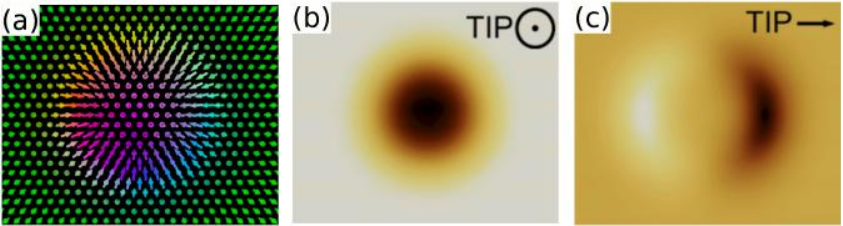}
\end{tabular}
\caption{\label{Fig1} (a) Spin structure of a skyrmion obtained from Ref.\ \onlinecite{rozsa-sk3}, and its constant-current SP-STM images \cite{palotas17prb} using (b) an out-of-plane and (c) an in-plane magnetized tip (bright: higher, dark: lower apparent height) according to Eq.\ (\ref{Eq_1a}).}
\end{figure}

To utilize the above electron charge and spin tunneling theory of noncollinear magnetic surfaces, we consider a magnetic skyrmion. Figure \ref{Fig1} reports the spin structure and SP-STM images of the charge current above the skyrmion showing characteristic circular and two-lobe contrasts for out-of-plane and in-plane magnetized tips, respectively \cite{romming15prl,palotas17prb}. It is known \cite{wortmann01,heinze06} that the charge current, $I\propto1+P_SP_T\cos\phi_A$, is sensitive to the effective spin polarization, $P_{\textrm{eff}}=P_SP_T$, see Eq.~(\ref{Eq_1a}). Therefore, the charge current has maxima at $\cos\phi_A=\pm1$ ($A_I=\pm P_{\textrm{eff}}$) and minima at $\cos\phi_A=\mp1$ ($A_I=\mp P_{\textrm{eff}}$) for $\mathrm{sgn}(P_{\textrm{eff}})=\pm1$.

\begin{figure*}[t]
\begin{tabular}{c}
\includegraphics[width=1.0\textwidth,angle=0]{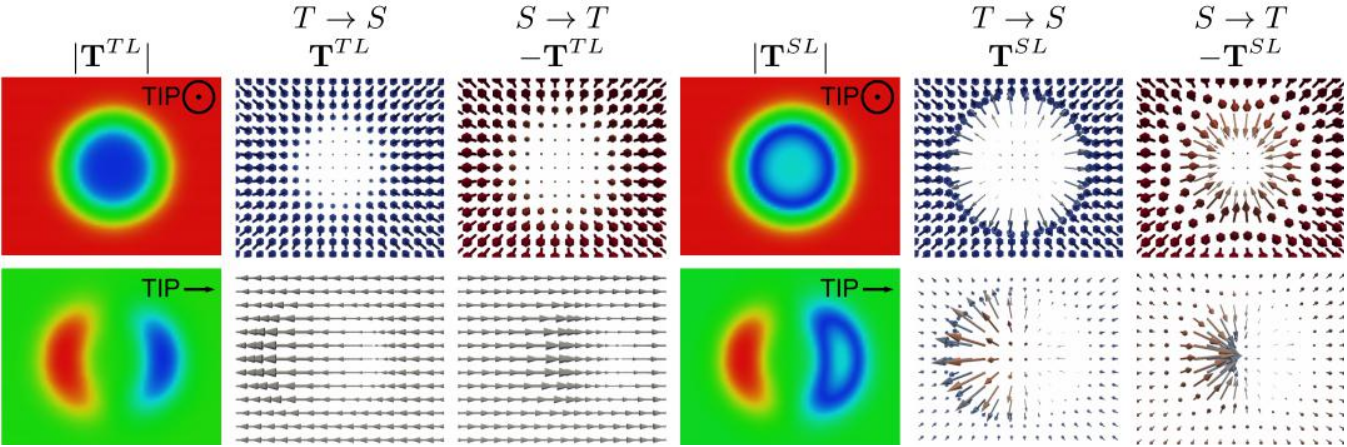}
\end{tabular}
\caption{\label{Fig2} Longitudinal spin current (LSC) magnitudes (red: maximum, blue: minimum) and vectors acting on the scanning tip ($|\mathbf{T}^{TL}|$ and $\mathbf{T}^{TL}$) and on the skyrmion ($|\mathbf{T}^{SL}|$ and $\mathbf{T}^{SL}$) for both $T\rightarrow S$ and $S\rightarrow T$ tunneling directions using an out-of-plane and an in-plane magnetized tip according to Eqs.\ (\ref{Eq_1b}) and (\ref{Eq_1c}). Red and blue colors of the LSC vectors correspond to positive and negative out-of-plane ($z$) vector components, respectively.}
\end{figure*}

Figure \ref{Fig2} shows calculated LSC magnitudes and vectors above the skyrmion in Fig.\ \ref{Fig1} with the same differently magnetized tips and the chosen spin polarization parameters. The maxima of the magnitudes $|\mathbf{T}^{TL}|$ and $|\mathbf{T}^{SL}|$ (red regions in Fig.\ \ref{Fig2}) are found at $\cos\phi_A=\pm1$ ($A_I=\pm P_{\textrm{eff}}$) for $\mathrm{sgn}(P_{\textrm{eff}})=\pm1$, exactly as for the charge current. However, the positions of the LSC minima (blue regions in Fig.\ \ref{Fig2}) depend on the relation of $P_S$ to $P_T$: if $|P_S|>|P_T|$ then $|\mathbf{T}^{TL}|$ have minima at $\cos\phi_A=-P_T/P_S$ ($A_I=-P_T^2$), and if $|P_S|<|P_T|$ then the minima are found at $\cos\phi_A=\mp1$ ($A_I=\mp P_{\textrm{eff}}$) for $\mathrm{sgn}(P_{\textrm{eff}})=\pm1$. Similarly, if $|P_T|>|P_S|$ then $|\mathbf{T}^{SL}|$ have minima at $\cos\phi_A=-P_S/P_T$ ($A_I=-P_S^2$), and if $|P_T|<|P_S|$ then the minima are found at $\cos\phi_A=\mp1$ ($A_I=\mp P_{\textrm{eff}}$) for $\mathrm{sgn}(P_{\textrm{eff}})=\pm1$. These result in an identical contrast of $|\mathbf{T}^{TL}|$ in Fig.\ \ref{Fig2} to that of the corresponding charge current in Fig.\ \ref{Fig1}. The contrast of $|\mathbf{T}^{SL}|$ is also qualitatively similar, except that its minima are found at $\cos\phi_A=-5/8$, $\phi_A=0.715\pi$ ($A_I=-0.25$), which are shown as blue belts in Fig.\ \ref{Fig2}. According to Eq.~(\ref{Eq_1b}), the $\mathbf{T}^{TL}$ vectors generally point to the $\mathrm{sgn}(P_T)\mathbf{s}_T$ direction, except for the regions with small LSC magnitudes if $|P_S|>|P_T|$. Similarly, according to Eq.~(\ref{Eq_1c}), the $\mathbf{T}^{SL}$ vectors generally point to the $\mathrm{sgn}(P_S)\mathbf{s}_S^A$ direction, except for the regions with small LSC magnitudes if $|P_T|>|P_S|$, that is inside the mentioned blue belts in Fig.\ \ref{Fig2}, where $\cos\phi_A<-5/8$. A detailed overview of the effect of various combinations of $P_S$ and $P_T$ on the LSC is given in Sec.\ \ref{sec_spinpol}.

The similarity of the LSC and the charge current image contrasts enables the estimation of the LSC based on experimentally measured SP-STM images. The LSC magnitudes can directly be related to the SP-STM images as discussed above, and the theoretical basis for this is outlined in Sec.\ \ref{sec_ilscstt}. The orientation of the LSC vectors can be based on the knowledge of $\mathbf{s}_T$ and the noncollinear spin structure $\mathbf{s}_S^a$. The latter can, in principle, be extracted from a series of SP-STM images with different tip magnetization directions \cite{palotas17prb,hagemeister16}, and such a procedure has been proven experimentally \cite{hsu17elec}. For the estimation of the LSC magnitudes and vectors, the knowledge of Eqs.~(\ref{Eq_1b})-(\ref{Eq_1c}) and the spin polarizations $P_S$ and $P_T$ are essential.

\begin{figure*}[t]
\begin{tabular}{c}
\includegraphics[width=1.0\textwidth,angle=0]{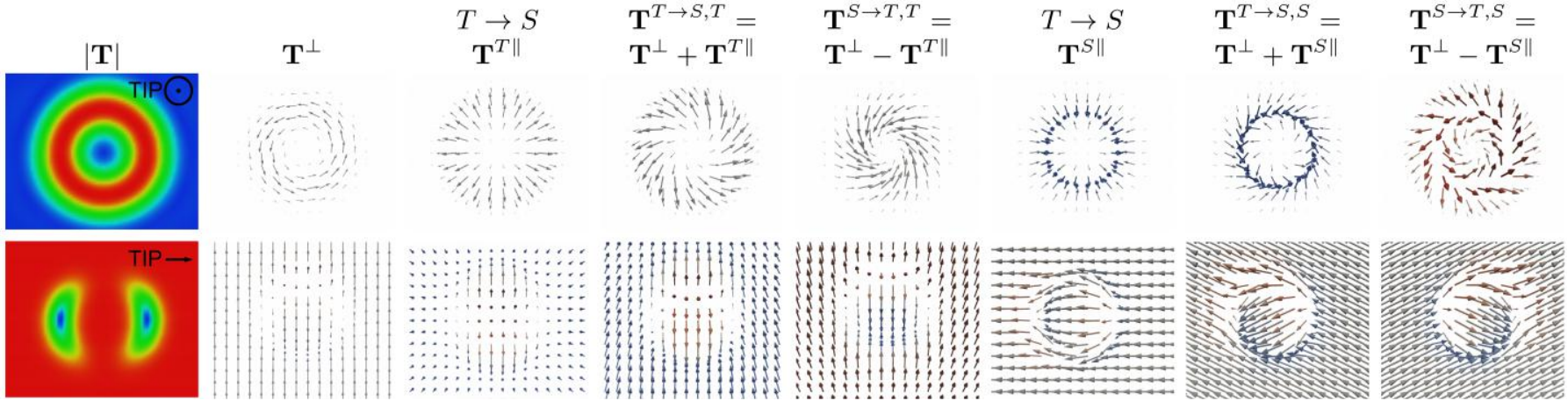}
\end{tabular}
\caption{\label{Fig3} Spin transfer torque (STT) magnitudes $|\mathbf{T}|$ (red: maximum, blue: minimum) and vectors (out-of-plane component ($\mathbf{T}^{\perp}$, Eq.\ (\ref{Eq_1f})), in-plane component ($\mathbf{T}^{j\parallel}$, Eqs.\ (\ref{Eq_1d}) and (\ref{Eq_1e})), total ($\mathbf{T}^{\perp}\pm\mathbf{T}^{j\parallel}$), depending on the tunneling direction $T\rightarrow S$ or $S\rightarrow T$, see Eqs.~(\ref{Eq_2a})-(\ref{Eq_2d})) acting on the spin moments of the scanning tip ($j=T$) and of the skyrmion ($j=S$) using an out-of-plane and an in-plane magnetized tip. Red and blue colors of the STT vectors correspond to positive and negative out-of-plane ($z$) vector components, respectively.}
\end{figure*}

Figure \ref{Fig3} shows calculated STT magnitudes, out-of-plane and in-plane STT vector components, and total STT vectors above the skyrmion in Fig.\ \ref{Fig1}. We find that the magnitudes of all STT components show the same type of contrast, which is denoted by $|\mathbf{T}|$ in Fig.\ \ref{Fig3}. Such a behavior results from their dominating $|\sin\phi_A|$-dependence due to the vector product $\mathbf{s}_S^A\times\mathbf{s}_T$ in Eqs.~(\ref{Eq_vectors4})-(\ref{Eq_vectors6}), with different spin-polarization-related prefactors (Eqs.\ (\ref{Eq_magnitudes3})-(\ref{Eq_magnitudes7})). Thus, the STT minima and maxima are obtained where the spins of the skyrmion are in line (parallel or antiparallel) with and perpendicular to the tip magnetization direction, respectively. Moreover, the STT minima at $\sin\phi_A=0$ are found exactly at the maxima and minima of the charge current, where $\cos\phi_A=\pm1$ ($A_I=\pm P_{\textrm{eff}}$), and the STT maxima are obtained at $\sin\phi_A=\pm1$ ($\cos\phi_A=0$, $A_I=0$). This means that the STT is small (large) where the absolute magnetic contrast of the charge current $|A_I|$ is large (small) \cite{palotas16prb}, see also Eqs.~(\ref{Eq_A4})-(\ref{Eq_A8}). This enables the estimation of the STT based on experimentally measured SP-STM images, similarly to the LSC, and again the knowledge of $\mathbf{s}_S^a$, $\mathbf{s}_T$, $P_S$, and $P_T$ is required, see the torque expressions in Eqs.~(\ref{Eq_1d})-(\ref{Eq_1f}) and (\ref{Eq_2a})-(\ref{Eq_2d}). The dependence of the STT on the combinations of $P_S$ and $P_T$ is investigated in Sec.\ \ref{sec_spinpol}.

The calculated STT vector components and vectors in Fig.\ \ref{Fig3} show a wide variety depending on the tip magnetization orientation ($\mathbf{s}_T$), the spin moment they are acting on ($T$ or $S$), and the tunneling direction ($T\rightarrow S$ or $S\rightarrow T$). For the out-of-plane magnetized tip ($\mathbf{s}_T\parallel z$, first row of Fig.\ 3) the $\mathbf{T}^{\perp}$ and $\mathbf{T}^{T\parallel}$ vectors are perpendicular to $z$, i.e., they lie in the $xy$ surface plane. The $\mathbf{T}^{T\parallel}$ vectors point to the direction of $P_S\mathbf{s}_S^A$ projected on the surface plane. Thus, the $\mathbf{T}^{\perp}\pm\mathbf{T}^{T\parallel}$ vectors are also in the surface plane. On the other hand, the $\mathbf{T}^{S\parallel}$ and the $\mathbf{T}^{\perp}\pm\mathbf{T}^{S\parallel}$ vectors have $z$-components proportional to $P_T$ (Eqs.~(\ref{Eq_1e})-(\ref{Eq_1f})), these are negative for $\mathbf{T}^{S\parallel}$ and $\mathbf{T}^{\perp}+\mathbf{T}^{S\parallel}$ and positive for $\mathbf{T}^{\perp}-\mathbf{T}^{S\parallel}$ in Fig.\ \ref{Fig3}. This difference in the $z$ component of the total STT vectors acting on the sample for the two tunneling directions has an important consequence for the possible rotation of the spins of the skyrmion due to the tunneling STT, clearly preferring one direction. For the skyrmion depicted in Fig.\ \ref{Fig1}(a) and the selected spin polarization parameters we conclude that $S\rightarrow T$ tunneling tends to annihilate the skyrmion since here the torque would rotate the spins outwards from the surface, while a spin rotation towards the surface would stabilize skyrmions in the case of $T\rightarrow S$ tunneling.

For the in-plane magnetized tip ($\mathbf{s}_T\parallel x$, second row of Fig.\ 3) the $\mathbf{T}^{\perp}$, $\mathbf{T}^{S\parallel}$ and $\mathbf{T}^{\perp}\pm\mathbf{T}^{S\parallel}$ vectors lie in the $xy$ surface plane outside the skyrmion above the ferromagnetic (FM) background, where $\mathbf{s}_S^A\parallel z$. Here, the $\mathbf{T}^{\perp}$ vectors are obtained as $z\times x=y$ and their direction ($\pm y$) is determined by the sign of $P_SP_T$. The $\mathbf{T}^{S\parallel}$ vectors are in line with $\mathbf{s}_T$ above the FM background, and their direction ($\pm x$) is determined by the sign of $P_T$. Here, the $\mathbf{T}^{T\parallel}$ vectors are in line with $\mathbf{s}_S^A$, and their direction ($\pm z$) is determined by the sign of $P_S$. By summing up the components, the total STT vectors $\mathbf{T}^{\perp}\pm\mathbf{T}^{T\parallel}$ and $\mathbf{T}^{\perp}\pm\mathbf{T}^{S\parallel}$ above the FM background can be characterized by two angles $\alpha_T$ and $\alpha_S$, which describe the inclination from the $\mathbf{s}_S^A\parallel z$ and $\mathbf{s}_T\parallel x$ directions, respectively. These angles are directly related to the spin polarizations of the tip and the sample,
\begin{subequations}
\begin{align}
|\tan\alpha_T|&=\frac{|\mathbf{T}^{\perp}|}{|\mathbf{T}^{T\parallel}|}=|P_T|,\label{Eq_alphaT}\\
|\tan\alpha_S|&=\frac{|\mathbf{T}^{\perp}|}{|\mathbf{T}^{S\parallel}|}=|P_S|.\label{Eq_alphaS}
\end{align}
\end{subequations}
Given the used spin polarization parameters in Fig.\ \ref{Fig3}, the two angles are $\alpha_T=\pm0.215\pi$ and $\alpha_S=\pm0.148\pi$. The corresponding inclinations of the total STT vectors above the FM background are clearly visible in the second row of Fig.\ \ref{Fig3}. Note that $-\pi/4\le\alpha_T,\alpha_S\le\pi/4$ since $-1\le P_T,P_S\le1$, and according to Eqs.\ (\ref{Eq_alphaT}) and (\ref{Eq_alphaS}) this means that the magnitude of $\mathbf{T}^{\perp}$ cannot exceed the magnitude of $\mathbf{T}^{\parallel}$ considering purely current-induced torques. Similarly, it was found in experiments performed for planar MTJs \cite{sankey08} that the magnitude of the out-of-plane torque is smaller than that of the in-plane torque. Equations~(\ref{Eq_alphaT}) and (\ref{Eq_alphaS}) also imply that the direct measurement of the STT vector components in magnetic tunnel junctions would allow an accurate determination of the spin polarizations of the tip and the sample separately. Presently, $P_{\textrm{eff}}=P_SP_T$ can be obtained from the measured charge current contrasts in the differential magnetic mode \cite{wulfhekel10} of SP-STM (see also Sec.\ \ref{sec_ilscstt}), and the knowledge of the spin polarization of one side is needed to determine the spin polarization of the other side in the tunnel junction. Further implications for STT measurements are given in Appendix.

\subsection{Effect of the spin polarizations on the tunneling charge and spin transport of a skyrmion}
\label{sec_spinpol}

\begin{figure*}[t]
\begin{tabular}{c}
\includegraphics[width=1.0\textwidth,angle=0]{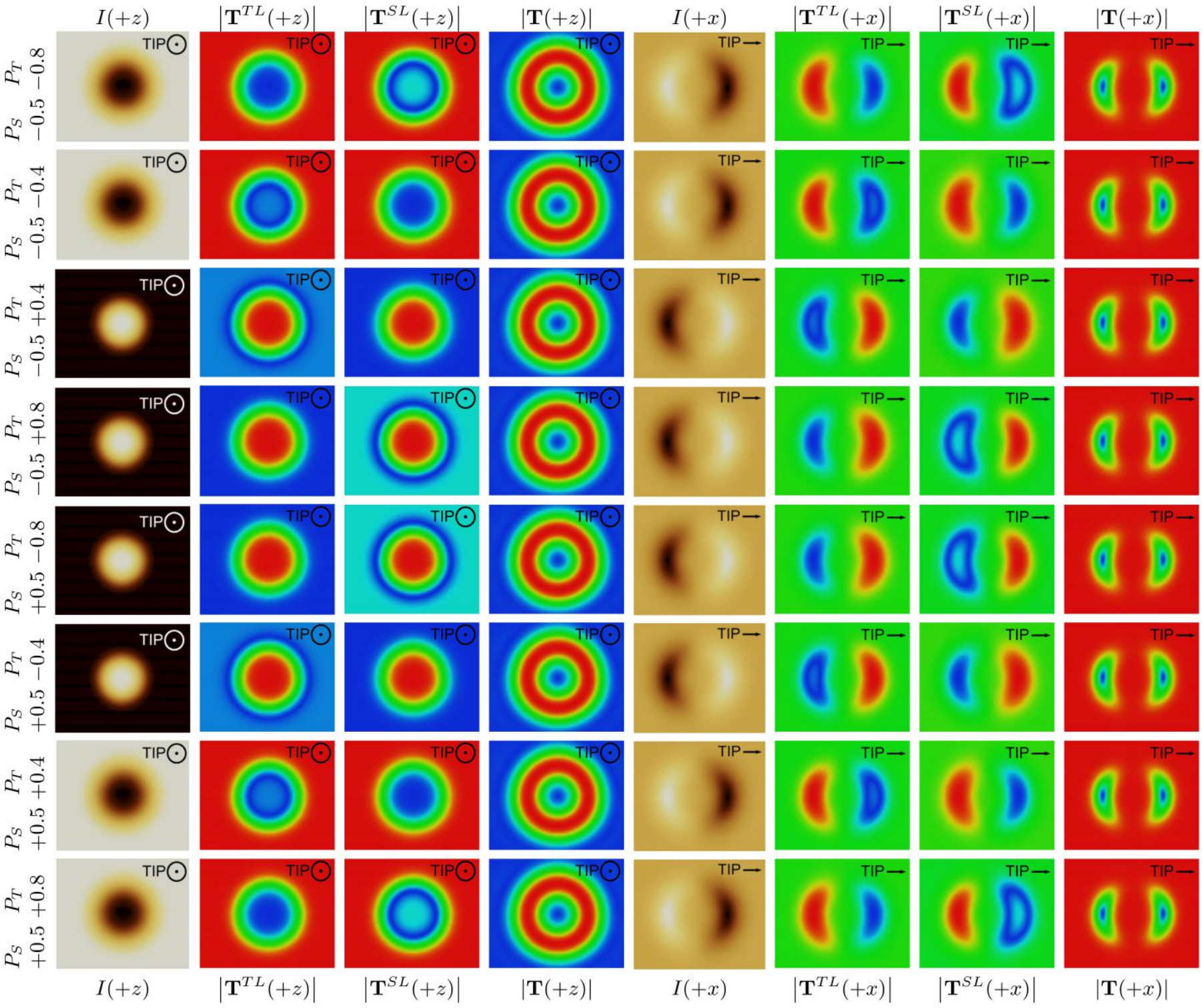}
\end{tabular}
\caption{\label{Fig4} Dependence of the charge and spin transport magnitudes on the spin polarizations of the surface ($P_S$) and the tip ($P_T$) in various combinations, for the skyrmion displayed in Fig.\ \ref{Fig1}: Constant-current SP-STM images ($I$; bright: higher, dark: lower apparent height), and magnitudes of the longitudinal spin current ($\left|\mathbf{T}^{TL}\right|$ and $\left|\mathbf{T}^{SL}\right|$; red: maximum, blue: minimum) and the spin transfer torque ($|\mathbf{T}|$; red: maximum, blue: minimum) at 6 \AA\;tip-sample distance using an out-of-plane ($+z$, in $[111]$ crystallographic direction) and an in-plane ($+x$, in $[1\bar{1}0]$ crystallographic direction) magnetized tip. The tip magnetization directions are explicitly shown. The color scales correspond to the data range of the individual images. Note that qualitatively very similar contrasts are observed for the magnitudes of the STT vectors and their components, i.e., for $|\mathbf{T}|$, $|\mathbf{T}^{\parallel}|$, and $|\mathbf{T}^{\perp}|$.}
\end{figure*}

\begin{figure*}[t]
\begin{tabular}{c}
\includegraphics[width=1.0\textwidth,angle=0]{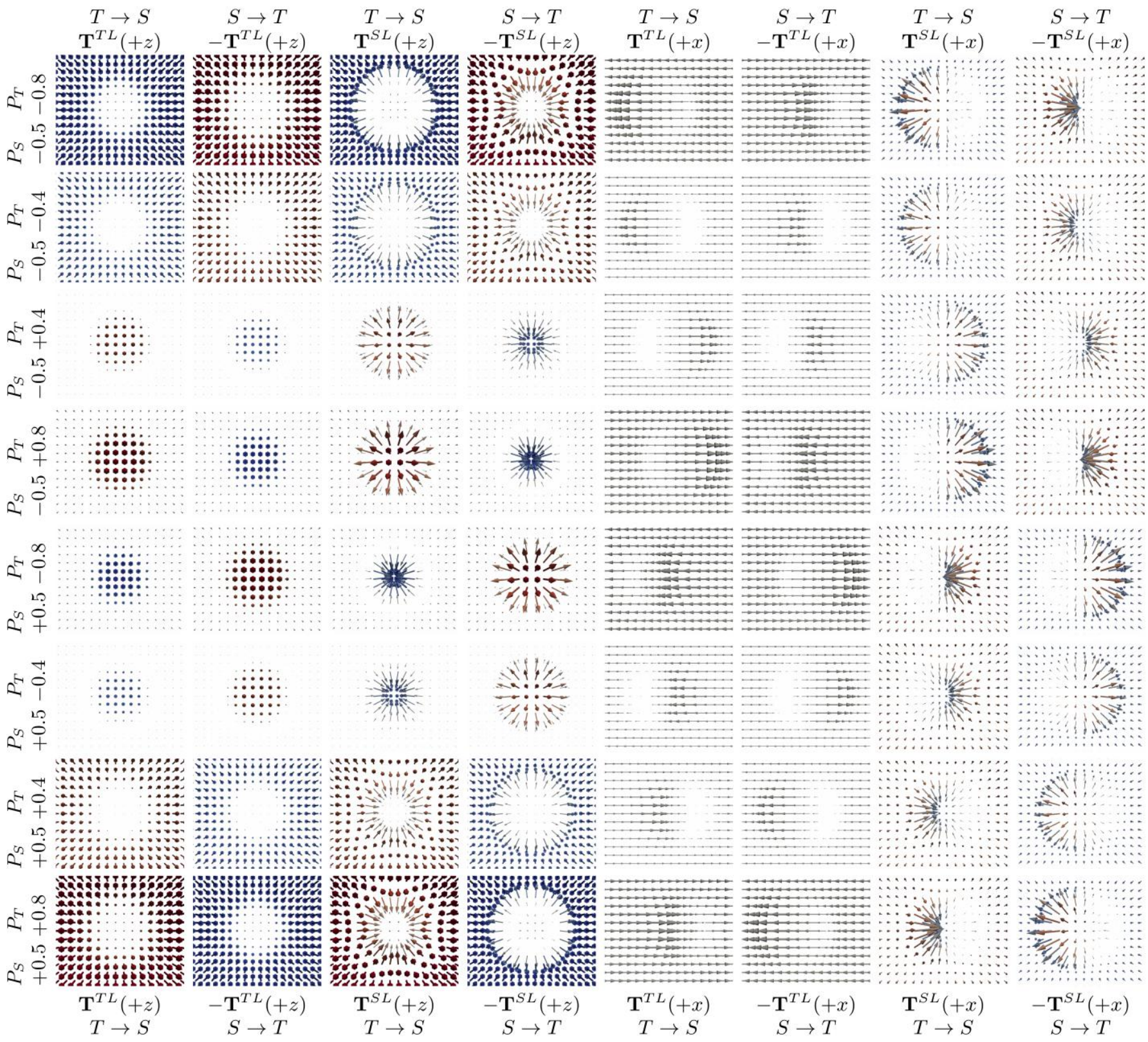}
\end{tabular}
\caption{\label{Fig5} Dependence of the longitudinal spin current (LSC) vectors ($\mathbf{T}^{TL}$ and $\mathbf{T}^{SL}$) on the spin polarizations of the surface ($P_S$) and the tip ($P_T$) in various combinations for the skyrmion in Fig.\ \ref{Fig1}. The LSC vectors are reported at 6 \AA\;tip-sample distance for both $T\rightarrow S$ and $S\rightarrow T$ tunneling directions using an out-of-plane ($+z$, in $[111]$ crystallographic direction) and an in-plane ($+x$, in $[1\bar{1}0]$ crystallographic direction) magnetized tip. The tip magnetization directions are explicitly shown in parentheses. Red and blue colors of the reported vectors correspond to positive and negative out-of-plane ($z$) components, respectively. The absolute maximal LSC magnitudes are 5.7 neV.}
\end{figure*}

\begin{figure*}[t]
\begin{tabular}{c}
\includegraphics[width=1.0\textwidth,angle=0]{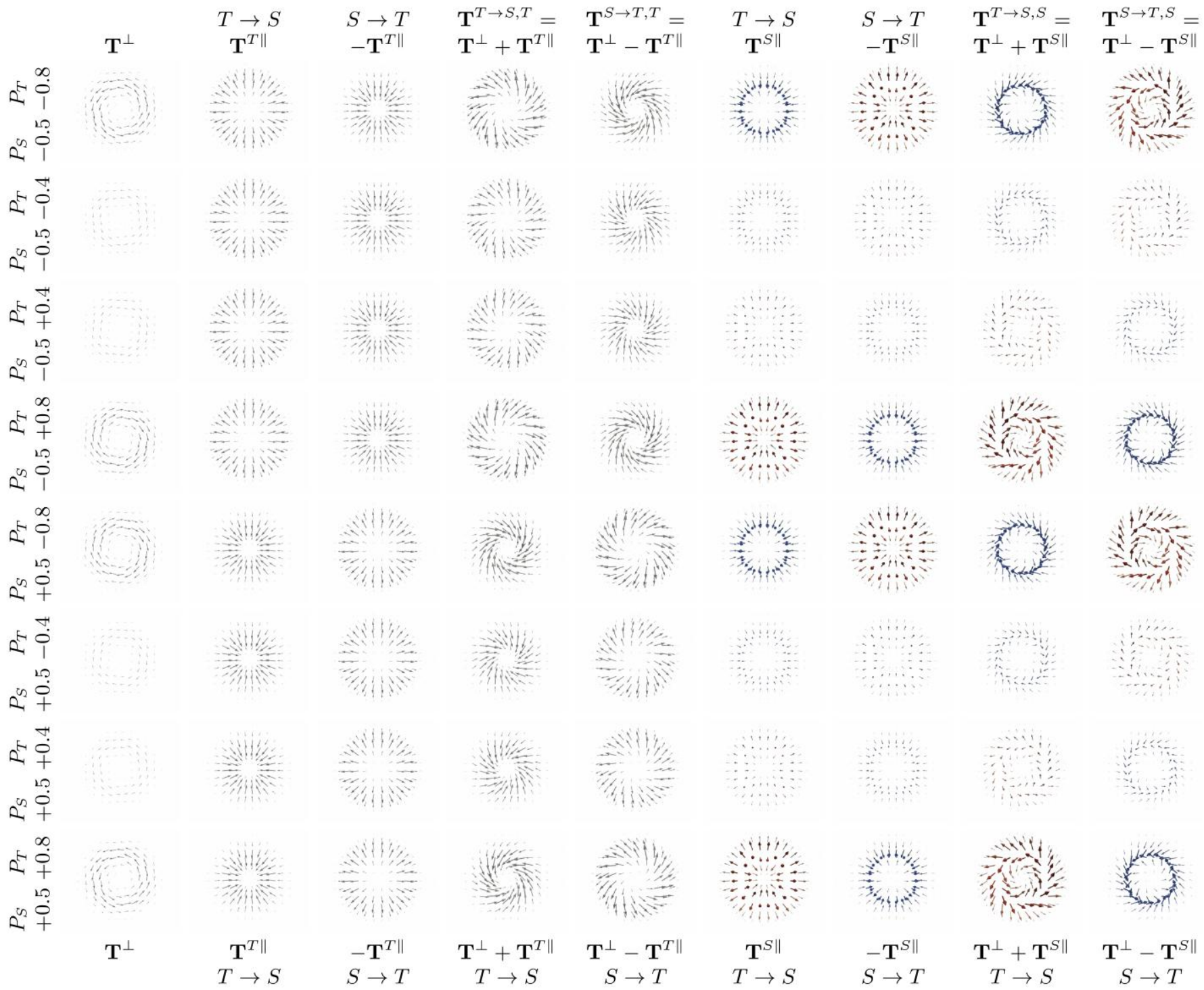}
\end{tabular}
\caption{\label{Fig6} Dependence of the spin transfer torque (STT) vector components and total STT vectors on the spin polarizations of the surface ($P_S$) and the tip ($P_T$) in various combinations for the skyrmion in Fig.\ \ref{Fig1} at 6 \AA\;tip-sample distance using an out-of-plane magnetized tip (spin moment pointing along the $+z$ or $[111]$ crystallographic direction). The STT vector components ($\mathbf{T}^{\perp}$, $\mathbf{T}^{j\parallel}$) and vectors ($\mathbf{T}^{\perp}\pm\mathbf{T}^{j\parallel}$) are acting on the spin moments of the scanning tip ($j=T$) and of the skyrmion ($j=S$). Red and blue colors of the reported vectors correspond to positive and negative out-of-plane ($z$) components, respectively. The absolute maximal STT magnitudes are 4 neV.}
\end{figure*}

\begin{figure*}[t]
\begin{tabular}{c}
\includegraphics[width=1.0\textwidth,angle=0]{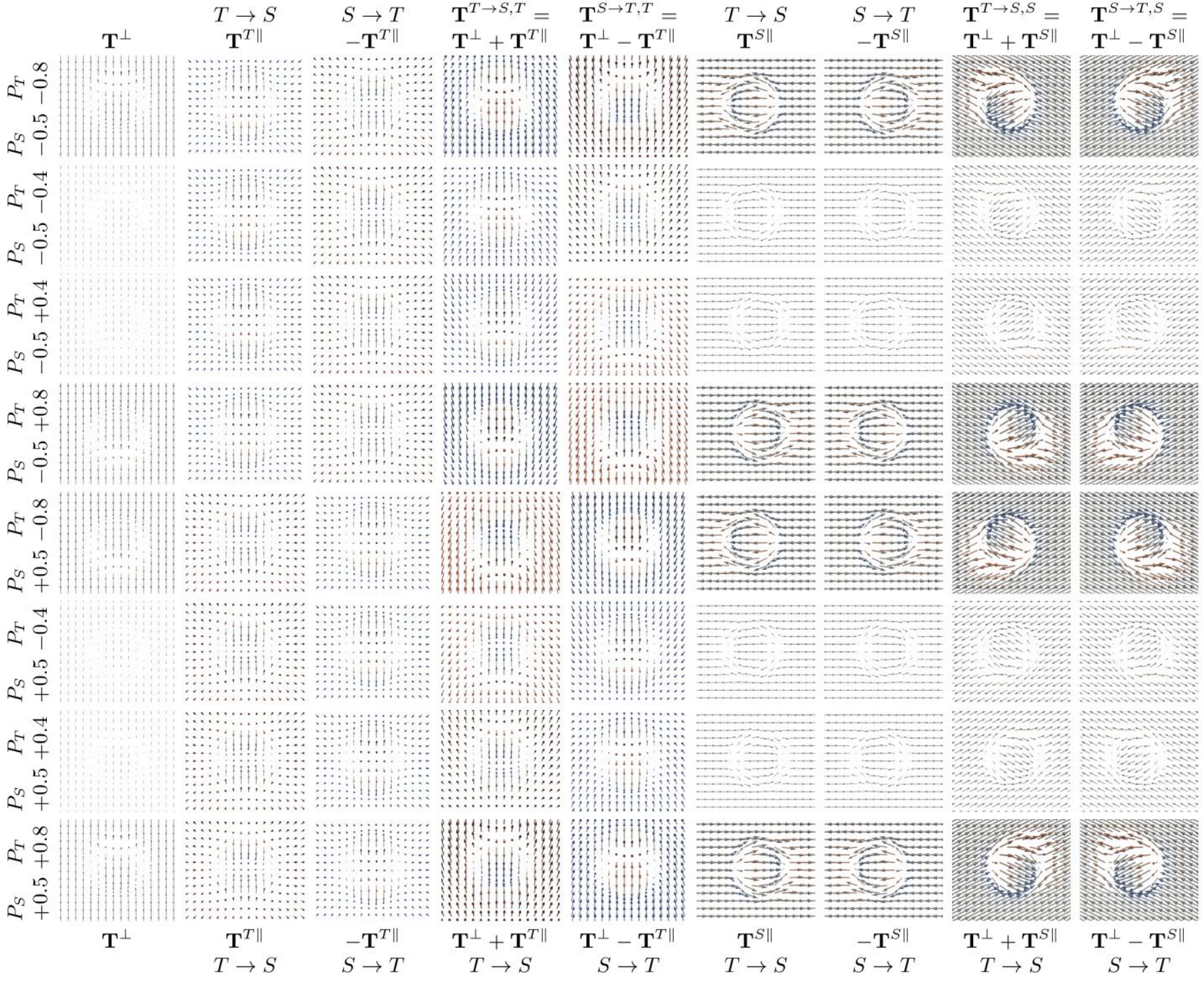}
\end{tabular}
\caption{\label{Fig7} Dependence of the spin transfer torque (STT) vector components and total STT vectors on the spin polarizations of the surface ($P_S$) and the tip ($P_T$) in various combinations for the skyrmion in Fig.\ \ref{Fig1} at 6 \AA\;tip-sample distance using an in-plane magnetized tip (spin moment pointing along the $+x$ or $[1\bar{1}0]$ crystallographic direction). The STT vector components ($\mathbf{T}^{\perp}$, $\mathbf{T}^{j\parallel}$) and vectors ($\mathbf{T}^{\perp}\pm\mathbf{T}^{j\parallel}$) are acting on the spin moments of the scanning tip ($j=T$) and of the skyrmion ($j=S$). Red and blue colors of the reported vectors correspond to positive and negative out-of-plane ($z$) components, respectively. The absolute maximal STT magnitudes are 4 neV.}
\end{figure*}

In the following, the tunneling charge and spin transport properties of the skyrmion in Fig.\ \ref{Fig1} are calculated and discussed taking the following combinations of the spin polarizations: $P_S\in\{-0.5,+0.5\}$ and $P_T\in\{-0.8,-0.4,+0.4,+0.8\}$.

Figure \ref{Fig4} displays SP-STM images of the charge current and the magnitudes of the spin transport quantities LSC and STT obtained with out-of-plane and in-plane magnetized tips, depending on $P_S$ and $P_T$. The SP-STM contrast is reversed by changing the sign of $P_{\textrm{eff}}$, see the contrasts of $I$ in the middle four images in the first and fifth columns of Fig.\ \ref{Fig4}. The LSC magnitudes in the second, third, sixth, and seventh columns of Fig.\ \ref{Fig4} show almost identical contrasts with those of the corresponding charge currents in the same row, and the contrast change depending on the sign of $P_{\textrm{eff}}$ is also reproduced. Such a behavior results from the expressions $|\mathbf{T}^{TL}|$ and $|\mathbf{T}^{SL}|$ in Eqs.~(\ref{Eq_magnitudes1}) and (\ref{Eq_magnitudes2}), respectively, and the direct connections between the LSC magnitudes and the charge current are introduced in Sec.\ \ref{sec_ilscstt}. The appearing blue belts for the $|\mathbf{T}^{TL}|$ and $|\mathbf{T}^{SL}|$ contrasts in Fig.\ \ref{Fig4} correspond to the real minima depending on the relation of $P_S$ to $P_T$, as explained at the discussion of Fig.\ \ref{Fig2} in Sec.\ \ref{sec_chsp}.

For the STT magnitudes we find the same type of contrast for both $|\mathbf{T}^{T}|$ and $|\mathbf{T}^{S}|$ and for all their components ($\perp$, $\parallel$, total), which are commonly denoted as $|\mathbf{T}|$ in the fourth and eighth columns of Fig.\ \ref{Fig4}. This is due to their dominating $|\sin\phi_A|$-dependence (Eqs.~(\ref{Eq_magnitudes3})-(\ref{Eq_magnitudes7})) as discussed at Fig.\ \ref{Fig3} in Sec.\ \ref{sec_chsp}. Figure \ref{Fig4} clearly shows that the contrasts of the STT magnitudes are sensitive to the magnetic structure only, and not to the involved spin polarizations. The STT minima (blue regions of STT in Fig.\ \ref{Fig4}) and maxima (red regions of STT in Fig.\ \ref{Fig4}) are obtained where the spins of the skyrmion are in line (parallel or antiparallel) with and perpendicular to the tip magnetization direction, respectively. Moreover, the STT minima are found exactly at the maxima and minima of the charge current, compare the corresponding $I$ and $|\mathbf{T}|$ contrasts in Fig.\ \ref{Fig4}.

Figure \ref{Fig5} shows calculated LSC vectors for the skyrmion in Fig.\ \ref{Fig1} for both $T\rightarrow S$ and $S\rightarrow T$ tunneling directions, depending on the considered combinations of $P_S$ and $P_T$, employing out-of-plane and in-plane magnetized tips. We find the general rule that $\mathbf{T}^{jL}(P_S,P_T)=-\mathbf{T}^{jL}(-P_S,-P_T)$, i.e., $\mathbf{T}^{T\rightarrow S,jL}(P_S,P_T)=\mathbf{T}^{S\rightarrow T,jL}(-P_S,-P_T)$ for $j\in\{T,S\}$. The magnitudes of the $\mathbf{T}^{TL}$ vectors correspond to the second and sixth columns of Fig.\ \ref{Fig4}, and the magnitudes of the $\mathbf{T}^{SL}$ vectors to the results shown in the third and seventh columns of Fig.\ \ref{Fig4}.

Figures \ref{Fig6} and \ref{Fig7} show calculated STT vectors and vector components for the skyrmion in Fig.\ \ref{Fig1} for both $T\rightarrow S$ and $S\rightarrow T$ tunneling directions, depending on the considered combinations of $P_S$ and $P_T$, employing an out-of-plane and an in-plane magnetized tip, respectively. We find that the $\mathbf{T}^{T\parallel}$ vectors scale with $P_S$, following its sign change, and are independent of $P_T$ (Eq.\ (\ref{Eq_1d})). Similarly, the $\mathbf{T}^{S\parallel}$ vectors scale with $P_T$, also following its sign change, and are independent of $P_S$ (Eq.\ (\ref{Eq_1e})). On the other hand, the $\mathbf{T}^{\perp}$ vectors scale with $P_SP_T$, and they follow the sign change of the effective spin polarization (Eq.\ (\ref{Eq_1f})). Since the total STT vectors are the sum of the corresponding two components, $\mathbf{T}^{j}=\mathbf{T}^{\perp}\pm\mathbf{T}^{j\parallel}$ with $j\in\{T,S\}$ for $T\rightarrow S$ and $S\rightarrow T$ tunneling directions, respectively, the STT results in Fig.\ \ref{Fig6} (with an out-of-plane magnetized tip) and in Fig.\ \ref{Fig7} (with an in-plane magnetized tip) give good indications on how the STT vectors can be tuned by changing the spin polarizations of the sample and the tip in the tunnel junction. This feature can turn out to be very useful if aiming at engineering the STT vectors at the atomic scale for technical applications in the future. We find the general rule that $\mathbf{T}^{T\rightarrow S,j}(P_S,P_T)=\mathbf{T}^{S\rightarrow T,j}(-P_S,-P_T)$ for $j\in\{T,S\}$. Note that the magnitudes of all STT components and vectors show the same type of contrast with a given tip magnetization orientation, and these contrasts are reported in the fourth and eighth columns of Fig.\ \ref{Fig4}, respectively. Taking an out-of-plane magnetized tip and the skyrmion depicted in Fig.\ \ref{Fig1}(a), we conclude that $S\rightarrow T$ tunneling tends to annihilate the skyrmion if $P_T<0$, and the opposite $T\rightarrow S$ tunneling direction tends to annihilate the skyrmion if $P_T>0$ because in both cases the $\mathbf{T}^S$ torque would rotate the spins outwards from the surface due to its positive $z$ component, see the last two columns of Fig.\ \ref{Fig6}.

Considering the different scalings of the STT vector components with $P_S$ (for $\mathbf{T}^{T\parallel}$), $P_T$ (for $\mathbf{T}^{S\parallel}$) or $P_SP_T$ (for $\mathbf{T}^{\perp}$), we can state that Eqs.~(\ref{Eq_alphaT}) and (\ref{Eq_alphaS}) generally hold true, and do not depend on the tip-sample geometry while in the tunneling regime. Deviations from this can be expected close to contact, where the importance of the STT contributions stemming from farther surface atoms below the STM tip apex is enhanced.

We find opposite inclinations of the STT vectors in the middle of the skyrmion compared to those above the ferromagnetic (FM) background in Fig.\ \ref{Fig7}. This is due to the opposite directions of the spins in that region compared to the FM background. Note that the determination of $P_S$ and $P_T$ from the inclinations of the total STT vectors, or from the in-plane and out-of-plane STT components refers to a certain bias voltage. In the presented model in Sec.\ \ref{sec_theory} we are restricted to very small bias and thus practically to the Fermi levels of both sides of the magnetic tunnel junction. Note, however, that the tunneling model can be extended to include energy dependence of the contributing electronic states, and bias voltage effects can be studied \cite{palotas16prb}. It is known that the bias voltage dependence of the charge current and the conductance complicates the determination of the energy-dependent spin polarizations significantly \cite{palotas12sts}. This is also expected in case of the determination of $P_S$ and $P_T$ from the STT vector components at non-zero bias voltage in possible future experiments.

\section{Summary and conclusions}
\label{sec_conc}

In summary, a theoretical method for the combined calculation of charge and vector spin transport of elastically tunneling electrons in high spatial resolution above complex noncollinear magnetic surfaces in SP-STM was developed. Connections between the SP-STM image contrasts of the charge current and the magnitudes of the longitudinal spin current (LSC) and the spin transfer torque (STT) were identified and explained. It was proposed that this enables the estimation of tunneling spin transport properties based on experimentally measured SP-STM images. A qualitative explanation was provided for a preferred bias voltage polarity for the STT contribution of skyrmion deletion in SP-STM. It was also proposed that the direct measurement of the STT vector components would enable the separate determination of the spin polarizations of the sample ($P_S$) and the tip ($P_T$), even above a ferromagnetic surface. A considerable tunability of the spin transport vectors by the involved spin polarizations was also demonstrated that could inspire the engineering of desired spin transport properties. The high-resolution determination of the tunneling STT and LSC vectors paves the way for the future investigation of current-induced magnetization switching in complex spin textures on surfaces due to local spin-polarized currents in SP-STM. As an example, the knowledge of the local STT and LSC vectors is expected to deliver a detailed microscopic insight into the creation, annihilation, and lateral manipulation of skyrmions and other complex surface magnetic objects in the future.

\section*{Acknowledgments}

Financial supports of the SASPRO Fellowship of the Slovak Academy of Sciences (project No.\ 1239/02/01), the Hungarian State E\"otv\"os Fellowship, the National Research, Development, and Innovation Office of Hungary under Projects No.\ K115575 and No.\ FK124100, the Alexander von Humboldt Foundation, and the Deutsche Forschungsgemeinschaft via SFB668 are gratefully acknowledged.

\appendix
\section{Spin transport vector measurement considerations}
\label{sec_measure}

Let us assume that in the magnetic STM junction the $\mathbf{T}^{\perp}$ and $\mathbf{T}^{T\parallel}$ vector components can be measured at low bias voltage for at least one tunneling direction $T\rightarrow S$ or $S\rightarrow T$, and denote $\mathbf{T}^{T\parallel}=\mathbf{T}^{T\rightarrow S,T\parallel}(=\mathbf{T}_s^{T\parallel}(V>0))=-\mathbf{T}^{S\rightarrow T,T\parallel}(=-\mathbf{T}_s^{T\parallel}(V<0))$, see Eq.~(\ref{Eq_torque_T1b}). With these the total STT vectors can be written as $\mathbf{T}^{T\rightarrow S,T}=\mathbf{T}^{\perp}+\mathbf{T}^{T\parallel}$ and $\mathbf{T}^{S\rightarrow T,T}=\mathbf{T}^{\perp}-\mathbf{T}^{T\parallel}$. Here, we show that by knowing the two STT vector components $\mathbf{T}^{\perp}$ and $\mathbf{T}^{T\parallel}$, or alternatively $\mathbf{T}^{T\rightarrow S,T}$ and $\mathbf{T}^{S\rightarrow T,T}$, the spin polarization of the tip ($P_T$) and the sample ($P_S$) and the magnitudes of the STT components and the STT acting on the sample surface can be determined. Taking the scalar product and using Eqs.~(\ref{Eq_magnitudes3}) and (\ref{Eq_magnitudes5}) result in
\begin{eqnarray}
\mathbf{T}^{T\rightarrow S,T}\cdot\mathbf{T}^{S\rightarrow T,T}&=&(\mathbf{T}^{\perp}+\mathbf{T}^{T\parallel})\cdot(\mathbf{T}^{\perp}-\mathbf{T}^{T\parallel})\nonumber\\
&=&|\mathbf{T}^{\perp}|^2-|\mathbf{T}^{T\parallel}|^2\nonumber\\
&=&P_S^2\sin^2\phi_A(P_T^2-1)\le0.
\end{eqnarray}
The absolute value square of the total STT vector is
\begin{eqnarray}
|\mathbf{T}^{T}|^2&=&|\mathbf{T}^{T\rightarrow S,T}|^2=|\mathbf{T}^{S\rightarrow T,T}|^2=|\mathbf{T}^{\perp}\pm\mathbf{T}^{T\parallel}|^2\nonumber\\
&=&|\mathbf{T}^{\perp}|^2+|\mathbf{T}^{T\parallel}|^2=P_S^2\sin^2\phi_A(P_T^2+1).
\end{eqnarray}
Taking the ratio of the above quantities results in a correlation-like formula,
\begin{eqnarray}
\rho_{\mathbf{T}^{T\rightarrow S,T},\mathbf{T}^{S\rightarrow T,T}}&=&\frac{\mathbf{T}^{T\rightarrow S,T}\cdot\mathbf{T}^{S\rightarrow T,T}}{\sqrt{\mathbf{T}^{T\rightarrow S,T}\cdot\mathbf{T}^{T\rightarrow S,T}}\sqrt{\mathbf{T}^{S\rightarrow T,T}\cdot\mathbf{T}^{S\rightarrow T,T}}}\nonumber\\
&=&-\frac{1-P_T^2}{1+P_T^2}\le0.
\end{eqnarray}
Using Eq.~(\ref{Eq_alphaT}), $P_T=\tan\alpha_T$,
\begin{equation}
\rho_{\mathbf{T}^{T\rightarrow S,T},\mathbf{T}^{S\rightarrow T,T}}=-\frac{1-\tan^2\alpha_T}{1+\tan^2\alpha_T}=-\cos(2\alpha_T).
\end{equation}
Thus, the angle $\alpha_T$ can be determined,
\begin{equation}
\alpha_T=\frac{1}{2}\arccos(-\rho_{\mathbf{T}^{T\rightarrow S,T},\mathbf{T}^{S\rightarrow T,T}}),
\end{equation}
and the following relationships can be observed,
\begin{eqnarray}
\frac{|\mathbf{T}^{\perp}|}{|\mathbf{T}^{T}|}&=&\frac{|P_T|}{\sqrt{1+P_T^2}}=|\sin\alpha_T|,\;\frac{P_T}{\sqrt{1+P_T^2}}=\sin\alpha_T,\nonumber\\
\frac{|\mathbf{T}^{T\parallel}|}{|\mathbf{T}^{T}|}&=&\frac{1}{\sqrt{1+P_T^2}}=\cos\alpha_T.
\end{eqnarray}
Knowing $P_T=\tan\alpha_T$, the magnitude of $\mathbf{T}^{S\parallel}$ is $|\mathbf{T}^{S\parallel}|=|P_T\sin\phi_A|$. Since we also know the out-of-plane STT vector $\mathbf{T}^{\perp}$, the same procedure can be repeated as above to determine $P_S=\tan\alpha_S$,
\begin{eqnarray}
\rho_{\mathbf{T}^{T\rightarrow S,S},\mathbf{T}^{S\rightarrow T,S}}&=&\frac{\mathbf{T}^{T\rightarrow S,S}\cdot\mathbf{T}^{S\rightarrow T,S}}{\sqrt{\mathbf{T}^{T\rightarrow S,S}\cdot\mathbf{T}^{T\rightarrow S,S}}\sqrt{\mathbf{T}^{S\rightarrow T,S}\cdot\mathbf{T}^{S\rightarrow T,S}}}\nonumber\\
&=&\frac{|\mathbf{T}^{\perp}|^2-|\mathbf{T}^{S\parallel}|^2}{|\mathbf{T}^{\perp}|^2+|\mathbf{T}^{S\parallel}|^2}\nonumber\\
&=&-\frac{1-P_S^2}{1+P_S^2}=-\frac{1-\tan^2\alpha_S}{1+\tan^2\alpha_S}=-\cos(2\alpha_S).\nonumber\\
\end{eqnarray}
Thus, the angle $\alpha_S$ can be obtained as
\begin{equation}
\alpha_S=\frac{1}{2}\arccos(-\rho_{\mathbf{T}^{T\rightarrow S,S},\mathbf{T}^{S\rightarrow T,S}}),
\end{equation}
and the following formulas relate the STT components to the STT magnitude,
\begin{eqnarray}
\frac{|\mathbf{T}^{\perp}|}{|\mathbf{T}^{S}|}&=&\frac{|P_S|}{\sqrt{1+P_S^2}}=|\sin\alpha_S|,\;\frac{P_S}{\sqrt{1+P_S^2}}=\sin\alpha_S,\nonumber\\
\frac{|\mathbf{T}^{S\parallel}|}{|\mathbf{T}^{S}|}&=&\frac{1}{\sqrt{1+P_S^2}}=\cos\alpha_S.
\end{eqnarray}

Inspired by the magnetic asymmetry of the charge current $A_I$ in Eq.\ (\ref{Eq_AI}), similar quantities can be defined for the LSC. For that reason, let us assume that the LSC vectors can be measured at opposite tip magnetization directions $\mathbf{s}_T$ and $-\mathbf{s}_T$. This results in the LSC vectors $\mathbf{T}^{TL}(\mathbf{s}_T)\propto(P_{T}+P_{S}\cos\phi_A)\mathbf{s}_T$, $\mathbf{T}^{TL}(-\mathbf{s}_T)\propto(P_{T}-P_{S}\cos\phi_A)(-\mathbf{s}_T$), $\mathbf{T}^{SL}(\mathbf{s}_T)\propto(P_{S}+P_{T}\cos\phi_A)\mathbf{s}_S^A$, and $\mathbf{T}^{SL}(-\mathbf{s}_T)\propto(P_{S}-P_{T}\cos\phi_A)\mathbf{s}_S^A$, from which the following magnetic asymmetry expressions can be obtained:
\begin{eqnarray}
A_{TL}&=&\frac{|\mathbf{T}^{TL}(\mathbf{s}_T)-\mathbf{T}^{TL}(-\mathbf{s}_T)|}{|\mathbf{T}^{TL}(\mathbf{s}_T)+\mathbf{T}^{TL}(-\mathbf{s}_T)|}\nonumber\\
&=&\frac{|P_T|}{|P_S\cos\phi_A|}=\frac{P_T^2}{|A_I|}=\frac{A_{SL}}{\cos^2\phi_A},\nonumber\\
A_{SL}&=&\frac{|\mathbf{T}^{SL}(\mathbf{s}_T)-\mathbf{T}^{SL}(-\mathbf{s}_T)|}{|\mathbf{T}^{SL}(\mathbf{s}_T)+\mathbf{T}^{SL}(-\mathbf{s}_T)|}\nonumber\\
&=&\frac{|P_T\cos\phi_A|}{|P_S|}=\frac{|A_I|}{P_S^2}=A_{TL}\cos^2\phi_A,
\end{eqnarray}
and $A_{TL}\ge A_{SL}\ge0$. With these quantities the spin transport magnitudes assuming an $\mathbf{s}_T$ tip magnetization direction can be written as
\begin{eqnarray}
|\mathbf{T}^{TL}|&\propto&|P_{T}|(1+1/A_{TL}),\nonumber\\
|\mathbf{T}^{SL}|&\propto&|P_{S}|(1+A_{SL}),\nonumber\\
|\mathbf{T}^{T\parallel}|&\propto&|P_{S}|\sqrt{1-A_{SL}/A_{TL}},\nonumber\\
|\mathbf{T}^{S\parallel}|&\propto&|P_{T}|\sqrt{1-A_{SL}/A_{TL}},\nonumber\\
|\mathbf{T}^{\perp}|&\propto&|P_{S}P_{T}|\sqrt{1-A_{SL}/A_{TL}},\nonumber\\
|\mathbf{T}^{T}|&\propto&|P_{S}|\sqrt{1+P_T^2}\sqrt{1-A_{SL}/A_{TL}},\nonumber\\
|\mathbf{T}^{S}|&\propto&|P_{T}|\sqrt{1+P_S^2}\sqrt{1-A_{SL}/A_{TL}}.
\end{eqnarray}
These equations establish a direct connection between the magnitudes of the spin transport components and the LSC asymmetries. Using Eq.\ (\ref{Eq_AI}), the spin polarizations can be expressed by solely using the magnetic asymmetries $A_I$, $A_{TL}$ and $A_{SL}$ as
\begin{eqnarray}
P_T^2&=&|A_I|\cdot A_{TL},\nonumber\\
P_S^2&=&|A_I|/A_{SL},\nonumber\\
P_T^2P_S^2&=&A_I^2A_{TL}/A_{SL},\nonumber\\
P_T^2/P_S^2&=&A_{TL}A_{SL}.
\end{eqnarray}
This way, the spin transport magnitudes assuming an $\mathbf{s}_T$ tip magnetization direction can be written using the magnetic asymmetries:
\begin{eqnarray}
|\mathbf{T}^{TL}|&\propto&\sqrt{|A_I|/A_{TL}}(1+A_{TL}),\\
|\mathbf{T}^{SL}|&\propto&\sqrt{|A_I|/A_{SL}}(1+A_{SL}),\nonumber\\
|\mathbf{T}^{T\parallel}|&\propto&\sqrt{|A_I|/A_{SL}-|A_I|/A_{TL}},\nonumber\\
|\mathbf{T}^{S\parallel}|&\propto&\sqrt{|A_I|A_{TL}-|A_I|A_{SL}},\nonumber\\
|\mathbf{T}^{\perp}|&\propto&|A_I|\sqrt{A_{TL}/A_{SL}-1},\nonumber\\
|\mathbf{T}^{T}|&\propto&\sqrt{|A_I|/A_{SL}-|A_I|/A_{TL}+A_I^2A_{TL}/A_{SL}-A_I^2},\nonumber\\
|\mathbf{T}^{S}|&\propto&\sqrt{|A_I|A_{TL}-|A_I|A_{SL}+A_I^2A_{TL}/A_{SL}-A_I^2}.\nonumber
\end{eqnarray}
The connections between the ratios of the magnitudes of the STT vector components (or the angles $\alpha_T$ and $\alpha_S$) and the magnetic asymmetries read as follows:
\begin{eqnarray}
\frac{|\mathbf{T}^{\perp}|}{|\mathbf{T}^{T}|}&=&|\sin\alpha_T|=\sqrt{\frac{|A_I|A_{TL}}{1+|A_I|A_{TL}}},\nonumber\\
\frac{|\mathbf{T}^{T\parallel}|}{|\mathbf{T}^{T}|}&=&\cos\alpha_T=\frac{1}{\sqrt{1+|A_I|A_{TL}}},\nonumber\\
\frac{|\mathbf{T}^{\perp}|}{|\mathbf{T}^{S}|}&=&|\sin\alpha_S|=\sqrt{\frac{|A_I|/A_{SL}}{1+|A_I|/A_{SL}}},\nonumber\\
\frac{|\mathbf{T}^{S\parallel}|}{|\mathbf{T}^{S}|}&=&\cos\alpha_S=\frac{1}{\sqrt{1+|A_I|/A_{SL}}}.
\end{eqnarray}
Other important relations can be written that might prove to be useful in the evaluation of future STT experiments in magnetic STM junctions:
\begin{eqnarray}
|\cos\phi_A|&=&\frac{|A_I|}{|\mathbf{T}^{TL}|\cdot|\mathbf{T}^{SL}|}(1+1/A_{TL})(1+A_{SL})\nonumber\\
&=&\sqrt{\frac{A_{SL}}{A_{TL}}}=\frac{|P_T|}{|P_S|}\frac{1}{A_{TL}}=\frac{|P_S|}{|P_T|}A_{SL}=\frac{|A_I|}{|P_S|\cdot|P_T|},\nonumber\\
|\sin\phi_A|&=&\frac{|\mathbf{T}^{T\parallel}|\cdot|\mathbf{T}^{S\parallel}|}{|\mathbf{T}^{\perp}|}=\sqrt{1-\frac{A_{SL}}{A_{TL}}}=\sqrt{1-\frac{A_I^2}{P_S^2P_T^2}},\nonumber\\
\frac{|\mathbf{T}^{T}|}{|\sin\alpha_S|}&=&\frac{|\mathbf{T}^{S}|}{|\sin\alpha_T|}=\frac{|\sin\phi_A|}{|\cos\alpha_S|\cdot|\cos\alpha_T|}.
\end{eqnarray}

\end{document}